\documentclass[review]{elsarticle}

\usepackage{lineno,hyperref}
\modulolinenumbers[5]
\usepackage{amsmath}
\usepackage{upgreek}
\usepackage{amssymb}
\usepackage[font=small,labelfont=bf]{caption}
\usepackage[font=small,labelfont=bf,tableposition=top]{caption}
\DeclareCaptionLabelFormat{andtable}{#1~#2  \&  \tablename~\thetable}
\usepackage[table,x11names]{xcolor}
\usepackage{float}
\renewcommand{\d}{\mathrm{d}}
\newcommand{\R}{\mathbb{R}}
\DeclareMathOperator{\grad}{grad}
\let\div\undefined
\DeclareMathOperator{\div}{div}

\journal{Journal of TBD}

\begin{document}

\begin{frontmatter}

\title{Data-driven Discovery of Chemotactic Migration \\ of Bacteria via Machine Learning}

\author[jhu]{Yorgos M. Psarellis}
\author[sjsu]{Seungjoon Lee}
\author[princeton2]{Tapomoy Bhattacharjee}
\author[princeton]{\newline Sujit S. Datta}
\author[jhu]{Juan M. Bello-Rivas}
\author[jhu,jhu1,jhu2]{Ioannis G. Kevrekidis\corref{mycorrespondingauthor}}
\cortext[mycorrespondingauthor]{Corresponding author}
\ead{yannisk@jhu.edu}

\address[jhu]{Department of Chemical and Biomolecular Engineering, Johns Hopkins University}
\address[sjsu]{Department of Applied Data Science, San Jos{\'e} University}
\address[princeton2]{Andlinger Center for Energy and the Environment, Princeton University}
\address[princeton]{Department of Chemical and Biological Engineering, Princeton University}
\address[jhu1]{Department of Applied Mathematics and Statistics, Johns Hopkins University}
\address[jhu2]{Department of Medicine, Johns Hopkins University}

\begin{abstract}
{\em E. coli} chemotactic motion in the presence of a chemoattractant field has been extensively studied using wet laboratory experiments, stochastic computational models as well as partial differential equation-based models (PDEs). The most challenging step in bridging these approaches, is establishing a closed form of the so-called chemotactic term, which describes how bacteria bias their motion up chemonutrient concentration gradients, as a result of a cascade of biochemical processes. Data-driven models can be used to learn the entire evolution operator of the chemotactic PDEs (black box models), or, in a more targeted fashion, to learn just the chemotactic term (gray box models).
In this work, data-driven Machine Learning approaches for learning the underlying model PDEs are (a) validated through the use
of simulation data from established continuum models and (b) used to infer 
chemotactic PDEs from experimental data. 
Even when the data at hand are sparse (coarse in space and/or time), noisy (due to inherent stochasticity in measurements) or partial (e.g. lack of measurements of the associated  chemoattractant field), we can attempt to learn the right-hand-side of a closed PDE for an evolving bacterial density. 
In fact we show that data-driven PDEs including a short history of the bacterial density field (e.g.
in the form of higher-order in time PDEs in terms of the measurable bacterial density)
can be successful in predicting further bacterial density evolution, and even possibly
recovering estimates of the unmeasured chemonutrient field. The main tool in this effort is the
effective low-dimensionality of the dynamics (in the spirit of the Whitney and Takens embedding theorems). The resulting data-driven PDE can then be simulated  to reproduce/predict computational or experimental bacterial density profile data, and estimate the underlying (unmeasured) chemonutrient field evolution.

\end{abstract}



\end{frontmatter}


\section{Author Summary}
\label{SecAuthor}

Bacteria, such as E.\textit{coli}, direct their motion towards (or away from) important chemical signals in their environment, a phenomenon known as chemotaxis. In this work, we use Machine Learning to learn \textit{laws} of chemotactic motion in a porous medium and in the presence of a nutrient. We show how our algorithms can predict coordinated bacterial movement in the macroscale. We also demonstrate how such algorithms can adapt to incorporate system-specific knowledge, or compensate for missing information using, instead a short history of observations. Our approaches are validated first with computational data and then tested with real-world chemotaxis experiments.

\section{Introduction}
\label{SecIntro}

We study models of the phenomenon of chemotactic migration of bacteria, i.e. their ability to direct multicellular motion along chemical gradients. This phenomenon is central to environmental, medical and agricultural processes \cite{Bhattacharjee_2020}.

Chemotaxis can be studied at several (complementary) levels: extensive fundamental research focuses on understanding the cellular mechanisms behind sensing chemoattractants/chemorepellents and how they induce a motility bias \cite{Sourjik123}.
Another approach is to understand and simulate chemotaxis in the context of stochastic processes and Monte Carlo methods \cite{Setayeshgar_2005, Siettos_2010, Othmer_2002, psarellis2022}. 
It is also possible, at appropriate limits, to employ macroscopic PDEs to simulate the spatiotemporal evolution of bacterial density profiles in the presence of a chemoattractant/chemorepellent field. In 
the latter approach, the Keller-Segel model has been notably successful \cite{ KELLER1971225}:

\begin{equation} \label{KS}
    \frac{\partial{b}}{\partial{t}}= \nabla \cdot ( D\nabla b - b \chi(S) \nabla S), \nonumber
\end{equation}
where $b(x,t)$ denotes the bacterial density, $D$ the diffusion coefficient and $S$
is the spatial field of the chemoattractant/chemorepellent 
(here considered constant in time). This model explicitly describes cell motility through two terms: a diffusion term (usually isotropic) and a chemotactic term, which encapsulates the response of the bacteria in the presence of a chemoattractant/chemorepellent field. This response includes signal transduction dynamics and properties of cellular chemoreceptors. In this term, the function $\chi: \mathbb{R} \rightarrow \mathbb{R}$  can be tuned for different kinds of chemotactically relevant substances and their spatial profile. Most importantly, the sign of $\chi $ distinguishes chemoattractants vs. chemorepellents. In general, the dynamics of  $S$ are described by a second PDE (for the field $S(x,t)$) coupled with the one above (see, for example Section \ref{methods_pde}). In this work, we will deal only with chemoattractants (and specifically chemonutrients).

Despite the generality and applicability of the Keller-Segel model, the chemotactic term is in general intractable. 
For example, to model chemotactic motion of {\em Escherichia coli (E. coli)} in heterogeneous porous media, Bhattacharjee et al., have used an  extension of the Keller-Segel model \cite{Bhattacharjee_2020} (see Section \ref{methods_pde}).
In that model, bacteria bias their motion towards the chemonutrient which they can consume (see second PDE in Eq. \ref{eq:1}). In that process, and for initial conditions corresponding to the experimental data presented here, they exhibit a macroscopic coherent propagating ``bacterial wave".

In this work, we demonstrate a toolbox of Data Mining methodologies that help learn different forms of {\em the law} of macroscopic chemotactic partial differential equations, either from simulations or from experimental data. These methods essentially help
construct macroscopic surrogate models either (a) for the entire right-hand-side of the PDEs (black-box) or (b) for a partial selection of those terms that are analytically unavailable/intractable (gray-box model).

Understanding and predicting the behavior of such a complex system is always a challenge. When the single-agent dynamics are known (possibly from first principles), the system can be studied and simulated at the microscale. 
Macroscale behavior naturally arises from simulating sufficiently large agent ensembles. It is sometimes possible to derive macroscale partial differential equations from the dynamics of the individual agents \cite{Othmer_2014, psarellis2022}. Such PDE-level descriptions are particularly attractive, as we are usually interested in the evolution of only a few, important macroscopic variables, rather than of the behavior of each individual ``microscopic'' agent (i.e. each individual bacterium).
Importantly, one also needs to know which (and how many) macroscopic variables/observables are sufficient to usefully construct a closed macroscopic evolution equation (e.g. \cite{Lee_2019}). 

For systems of great complexity, an accurate macroscopic PDE may be out of reach. One could only gather (full or partial) information from experiments and/or fine-scale, individual based/stochastic simulations.  This calls for a data-driven approach to ``discover" a macroscopic law for a coarse PDE, solely from spatiotemporal data (experimental or computational movies). Such a data-driven PDE can then be exploited towards the following purposes:

\begin{enumerate}
   
    \item Predicting the time evolution starting from different initial conditions or in out-of-sample spatiotemporal domains. This is particularly attractive when it is not easy to probe the system and extract such profiles ``on demand'' from experiments or simulations.
    \item Reconstructing the full behavior of the system even when only partial information is at hand (i.e. when we do not have data for all important macroscopic variables). 
    \item If a qualitatively correct but quantitatively inaccurate macroscopic model is  available, a quantitative data-driven model can help probe and even understand different components of the system's behavior \cite{psarellis2022}. 
    This can be a way to shed light on the fundamental physical laws of the studied system (explainability). 
\end{enumerate}

Our work falls in the general category of nonlinear system identification using data-driven, machine-learning-assisted surrogate models. Neural Networks have repeatedly  demonstrated  successes in learning nonlinear Ordinary \cite{Rico-Martinez1994596} or Partial Differential Equations \cite{Lee_2019, GONZALEZGARCIA1998S965}. More recently, with the increased accessibility of powerful computational hardware and the computational efficiency of Machine Learning algorithms, nonlinear system identification has attracted a lot of attention \cite{Galaris2022} and has motivated the design of novel approaches and architectures. Notable among these  approaches are Neural ODEs \cite{chen2019neural} and Convolutional Neural Networks \cite{lecun1995convolutional, rao2022discovering}, sparse identification \cite{brunton2016} and effective dynamics identification \cite{vlachas2018, Vlachas2022}.


\section{Models and Results}
\label{SecMethods_results}



\noindent We constructed data-driven models for two different data sources (which, however correspond to the same general chemotaxis scenario):

\begin{enumerate}[(i)]
    \item PDE simulation data: The Chemotactic PDE described and used in \cite{Bhattacharjee_2020} was simulated and data were collected for both bacterial density and nutrient concentration fields. Details about this PDE and its simulation can be found in Section \ref{SecMethods_results}.
    \item Real-world data from chemotaxis experiments were used. These experiments were performed by Bhattacharjee et al. and described in \cite{Bhattacharjee_2020}. 
\end{enumerate}

The results of this Section are presented separately for each of these two categories.

\subsection{Models for Simulation Data}
\label{subsec:sim}

PDE simulations included the integration of two coupled PDE fields, one describing the bacterial density $b$ and one describing the nutrient density $c$ (System \ref{eq:1}). The PDE solution can be seen in Fig.\ref{fig:ground_truth}.

To learn from simulation data, we considered a variety of machine-learning-enabled data-driven models; they are listed in Table \ref{tab2:model_sum} and Table \ref{tab2:model_sum_c}; the relevant notation is summarized in the Table captions. In the text that follows, a representative selection (the highlighted models in Tables \ref{tab2:model_sum}, \ref{tab2:model_sum_c} ) will be described in more detail; the remaining ones are relegated to the Supplementary Information.

\vspace{0.8cm}

\begin{table}[H]
    \makebox[\textwidth][c]{

     \begin{tabular}{||c c c c c c||} 
     \hline
     Model & Surrogate Function & Known Fields & Known RHSs & Output & Algorithm \\ [0.5ex] 
     \hline\hline
     Black-box for 2 PDEs & $f_{GP}, h_{GP}$ & $b(r,t), c(r,t)$ & -- & $b_t, c_t$  & GPR\\
      \rowcolor{yellow} & $f_{NN}, h_{NN}$ &  $b(r,t), c(r,t)$ & -- & $b_t, c_t$  & ANN\\
    Black-box for 1 PDE& $f_{GP}$ &  $b(r,t), c(r,t)$ & $c_t$ & $b_t$  & GPR\\
      \rowcolor{yellow} & $f_{NN}$ &  $b(r,t), c(r,t)$ & $c_t$ & $b_t$  & ANN\\
    Black-box, delays& $f^{partial}_{GP}$ &   $b(r,t)$, history &  -- & $b(r, t+\Delta t)$  & GPR\\
        \rowcolor{yellow} & $f^{partial}_{NN}$&  $b(r,t)$, history & -- & $b(r, t+\Delta t)$  & ANN\\
    Gray-box& $g_{GP}$ &   $b(r,t), c(r,t)$ & $c_t$ & $b_t-D_b\Delta b$  & GPR\\
       \rowcolor{yellow} & $g_{NN}$ &  $b(r,t), c(r,t)$ & $c_t$ & $b_t-D_b\Delta b$  & ANN\\
    Gray-box, delays& $g^{partial}_{GP}$ &  $b(r,t)$, history & -- & $b(r, t+\Delta t)-D_b\Delta b(r,t)$  & GPR\\
      & $g^{partial}_{NN}$ &  $b(r,t)$, history & -- & $b(r, t+\Delta t)-D_b\Delta b(r,t)$  & ANN\\[1ex]
     \hline
    \end{tabular}}
\vspace{0.6cm}
\caption{Listing of all data-driven models explored in the manuscript. Notation: $f, h$ denote surrogate functions for the entire RHS of the $b-$ and  $c- $PDE respectively, while $g$ denotes a surrogate for the chemotactic term. Subscripts ``GPR" and ``NN" denote Gaussian Process Regression and Artificial Neural Network respectively. $\Delta t$ denotes the time delay used in all models with partial information. Only the highlighted models will be presented in detail in the following sections; the rest are included in the Supplementary Information.}
\label{tab2:model_sum}
\end{table}

\begin{center}
\begin{table}[h]
     \begin{tabular}{||c c c c c||} 
     \hline
     Surrogate Function & Known Fields & Known RHSs & Output & Algorithm \\ [0.5ex] 
     \hline\hline
     
      $C_{GP}$ &   $b(r,t)$, history & -- & $c(r,t)$  & GPR\\
      \rowcolor{yellow} $C_{NN}$ &  $b(r,t)$, history & -- & $c(r,t)$  & ANN\\
     
     \hline
    \end{tabular}
\vspace{0.2cm}
\caption{Summary of all data-driven models for learning the nutrient field (denoted as  $c(x,t)$). Only the highlighted model will be presented in detail in this section. The other model is included in the Supplementary Information.}
\label{tab2:model_sum_c}
\end{table}
\end{center}

\begin{center}
     \makebox[\textwidth][c]{\includegraphics[width=20cm, height=10cm]{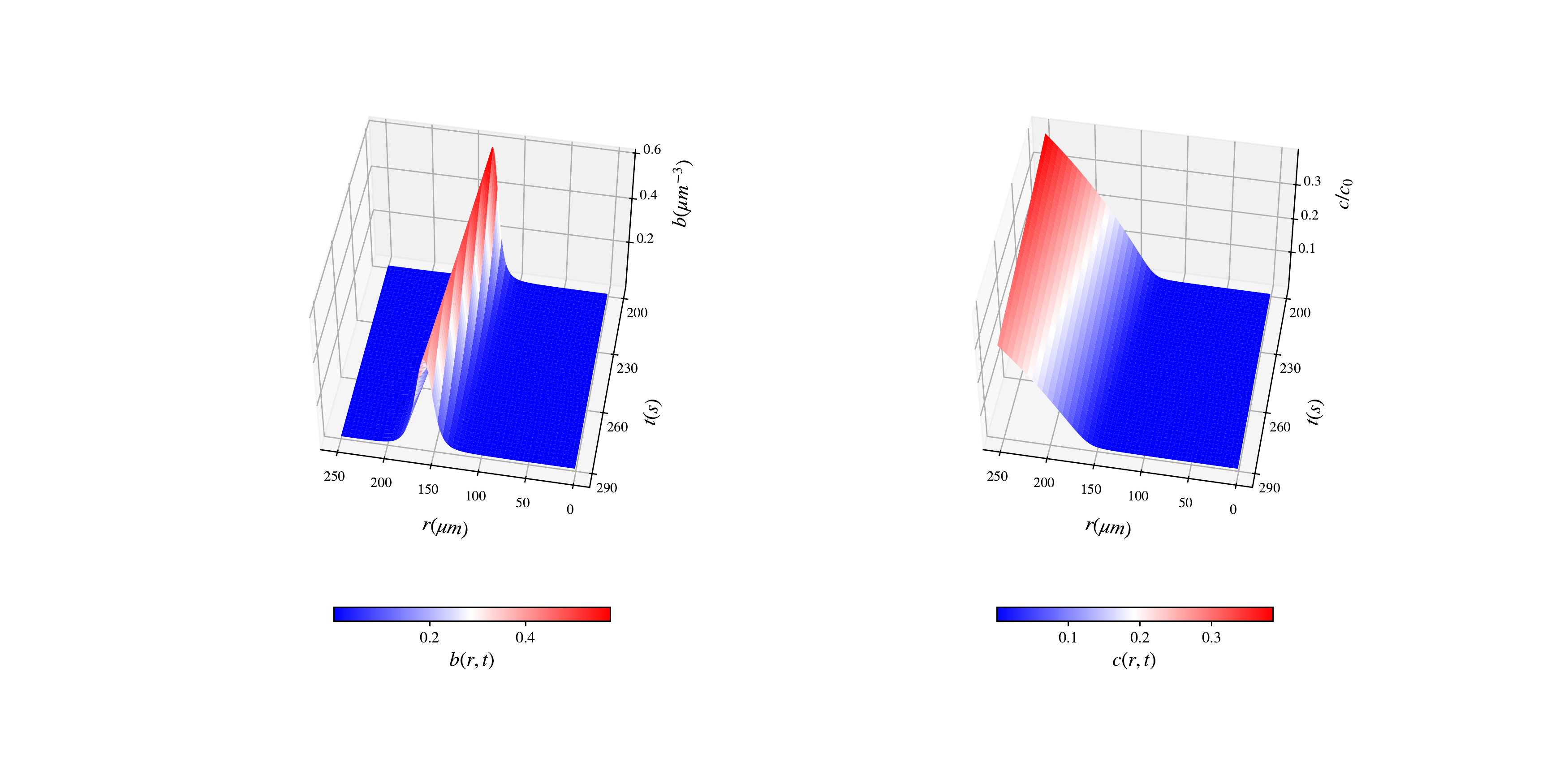}}
     \captionof{figure}{PDE simulations representing the ground truth of the simulations: (left) $b(r,t)$ field and (right) $c(r,t)$ field. For clarity, arrows are added to denote the direction of time.}
     \label{fig:ground_truth}
\end{center}

\subsubsection{Black box data-driven models}
\label{subsubsec:blackbox_sim}

Consider a system described by $d$ macroscopic scalar variable fields ($u^{(1)}, \\... , u^{(d)}$). Assuming a one-dimensional domain along the vector $\mathbf{\hat{x}}$ (for an example in cylindrical coordinates see Figure. \ref{figtab}) discretized through $m$ points in space ($x$) and $n$ points in time ($t$), we are given 
$m\cdot n$ data points in $\mathbb{R}^{d}$. Using interpolation/numerical differentiation, we can estimate the temporal derivatives $u^{(1)}_t,... , u^{(d)}_t$, as well as various order derivatives in space (first, $u^{(1)}_x,... , u^{(d)}_x$, second $u^{(1)}_{xx},... , u^{(d)}_{xx}$, etc). We assume that we know \textit{a priori} the largest order of relevant spatial derivatives (here, two) \cite{li2003}, the coordinate system, and the boundary conditions (here, zero flux). We also assume that the spatiotemporal discretization satisfies the necessary criteria for a numerically converged PDE solution. Given these derivatives, we can compute all relevant local operators, such as: $\mathbf{\nabla u^{(i)}}, \Delta u^{(i)},  i \in \{1, ..., d\}$.
In Cartesian coordinates these operators are simply related to the spatial derivatives; but in curvilinear coordinates, or when the evolution occurs on curved manifolds, the relation between spatial derivatives and local operators needs a little more care. 
We consider physical Euclidean space $\R^3$ (regarded as a Riemannian manifold with Euclidean metric expressed as $g = (\d x)^2 + (\d y)^2 + (\d z)^2$ in Cartesian coordinates $(x, y, z)$).
The gradient, $\grad f$, of a smooth function $f \colon \R^3 \to \R$ is a vector pointing in the direction at which $f$ grows at its maximal rate and whose length is said maximal rate ---note that this definition is independent of the system of coordinates on $\R^3$.
%
%
The phase flow of a smooth vector field $v \colon \R^3 \to \R^3$ can be regarded as the motion of a fluid in space.
The divergence of $v$, denoted by $\div v$, is the outflow of fluid per unit volume at a given point.
Again, the definition is coordinate frame independent;
the expression $\div v = \frac{\partial v_1}{\partial x} + \frac{\partial v_2}{\partial y} + \frac{\partial v_3}{\partial z}$ is valid in Cartesian coordinates.
The Laplacian is the composition of the gradient followed by the divergence (in other words, $\Delta = \div \grad$), again independently of the choice of coordinates.

Our objective is to learn functions $f_i: \mathbb{R}^{3d \cdot m} \rightarrow \mathbb{R}^m, i \in \{1, ..., d\}$ such that:

\begin{equation} \label{ddPDE_BB}
    u^{(i)}_t= f_i( u^{(1)}, ...,  u^{(d)},\mathbf{\nabla u^{(1)}} \cdot \mathbf{\hat{x}}, ..., \mathbf{\nabla u^{(d)}} \cdot \mathbf{\hat{x}},\Delta u^{(1)}, ...,  \Delta u^{(d)} ) \nonumber
\end{equation}

\noindent This is a \textbf{black box expression} for the time derivative of a macroscopic field
expressed as a function of the relevant lower order \textit{coordinate-independent} local spatial
operators, operating on the fields. After training (after successfully learning this function based on data)
integrating this model numerically can reproduce 
spatiotemporal profiles in the training set, and even hopefully predict them beyond the training set.
The arguments of $f_i$ will be the features (or input vectors) and $u^{(i)}_t$ will be the target (or output vector) of our data-driven model. Note that, usually, not all features are informative in the learning of $f_i$ (in other words, only some orders of the spatial derivatives appear in the PDE right-hand-side).  Also, note that not all macroscopic variables $u^{(i)}$ are always necessary for learning  $f_j, j \neq i$. 
In the spirit of the Whitney and Takens embedding theorems \cite{Whitney_1936, Takens_1981},
short histories of some of the relevant variable profiles may ``substitute" for missing
observations/measurements of other relevant variables.

Note that in our specific case of \textbf{cylindrical coordinates} with only the component along the radial dimension (with unit vector denoted $\mathbf{\hat{r}}$) being important due to domain-specific symmetries, the right-hand-side of a PDE will explicitly depend on the local radius as well.  In the above formulation, this is incorporated in the Laplacian term
$\Delta u^{(i)} =  \frac{1}{r}  \frac{\partial}{\partial r} (r  \frac{\partial  u^{(i)}}{\partial r} )$. For the construction of all relevant differential operators in any coordinate system, it is possible to use ideas and tools from Exterior Calculus (see SI).


\paragraph{Black box learning of both PDEs with an ANN (from known fields $b(r,t), c(r,t)$)} 
\label{SubSubSec2BBNN}

 \begin{align} \label{eq2NNBB}
   \begin{bmatrix}
           b_t \\
           c_t \\
           \end{bmatrix} &= \begin{bmatrix}
           f \\
           h \\
           \end{bmatrix} = \mathbf{F_{NN}}(b, \mathbf{\nabla b} \cdot \mathbf{\hat{r}},\Delta b, c, \mathbf{\nabla c} \cdot \mathbf{\hat{r}},\Delta c )
  \end{align}

 \begin{center} 
     \makebox[\textwidth][c]{\includegraphics[width=17.5cm, height=7cm]{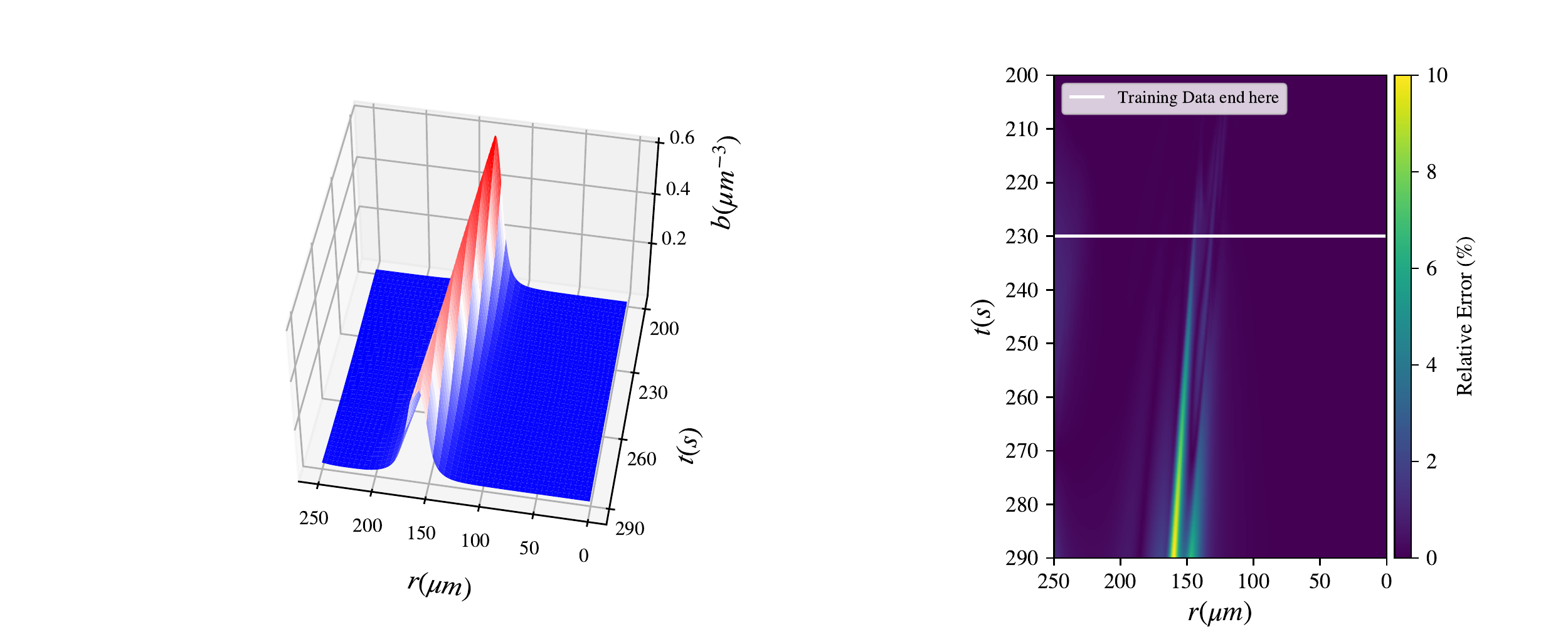}}
     \captionof{figure}{Black-box learning of both PDEs with an Artificial Neural Network: (left) Integration results for the \textbf{first} data-driven PDE  (for $b(r,t)$)   and (right) relative error (\%).}
\label{fig:fig2NNBBB}
\end{center}
\vspace{0.0mm}
\begin{center} 
     \makebox[\textwidth][c]{\includegraphics[width=17.5cm, height=7cm]{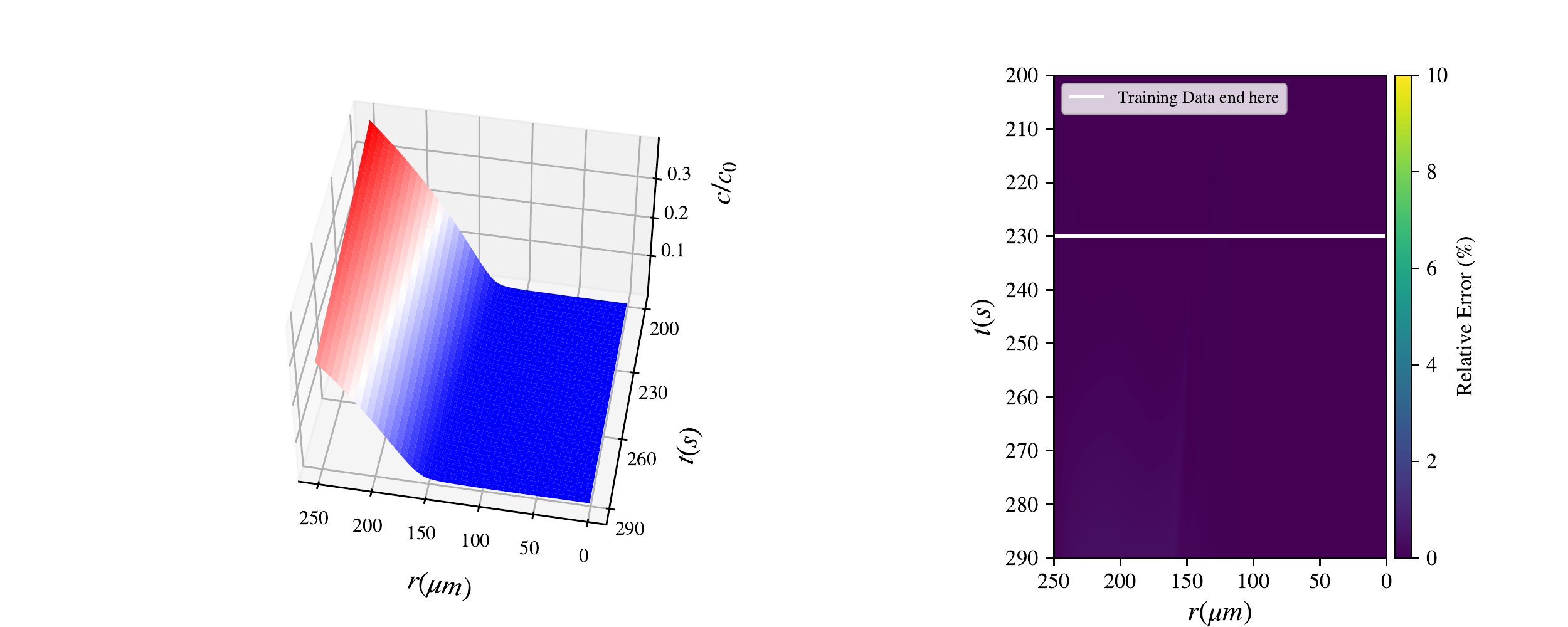}}
     \captionof{figure}{Black-box learning of both PDEs with an Artificial Neural Network: (left) Integration results for the \textbf{second} data-driven PDE (for $c(r,t)$) and (right) relative error (\%).}
\label{fig:fig2NNBBC}
\end{center}

With training data from both $b(r,t)$ and $c(r,t)$ fields, we learned a neural network of the form described in equation \ref{eq2NNBB}.

Figures \ref{fig:fig2NNBBB}, \ref{fig:fig2NNBBC} show how the data-driven PDEs were able to learn the laws of time evolution of both $b(r,t)$ and $c(r,t)$. The data-driven models were used to reproduce the training data and could successfully extrapolate as far as we attempted (here, up to $t=290\mathrm{s}$).

\vspace{1cm}

\paragraph{Black box learning of $b_t$ with ANN - $c_t$  known (with fields $b(r,t), c(r,t)$ known)}
\label{SubSubSecBBNN}

\begin{equation} \label{eqNNBB}
    b_t= f_{NN}(b, \mathbf{\nabla b} \cdot \mathbf{\hat{r}},\Delta b, c, \mathbf{\nabla c} \cdot \mathbf{\hat{r}},\Delta c )
\end{equation}

\begin{center}
     \makebox[\textwidth][c]{\includegraphics[width=20cm, height=8cm]{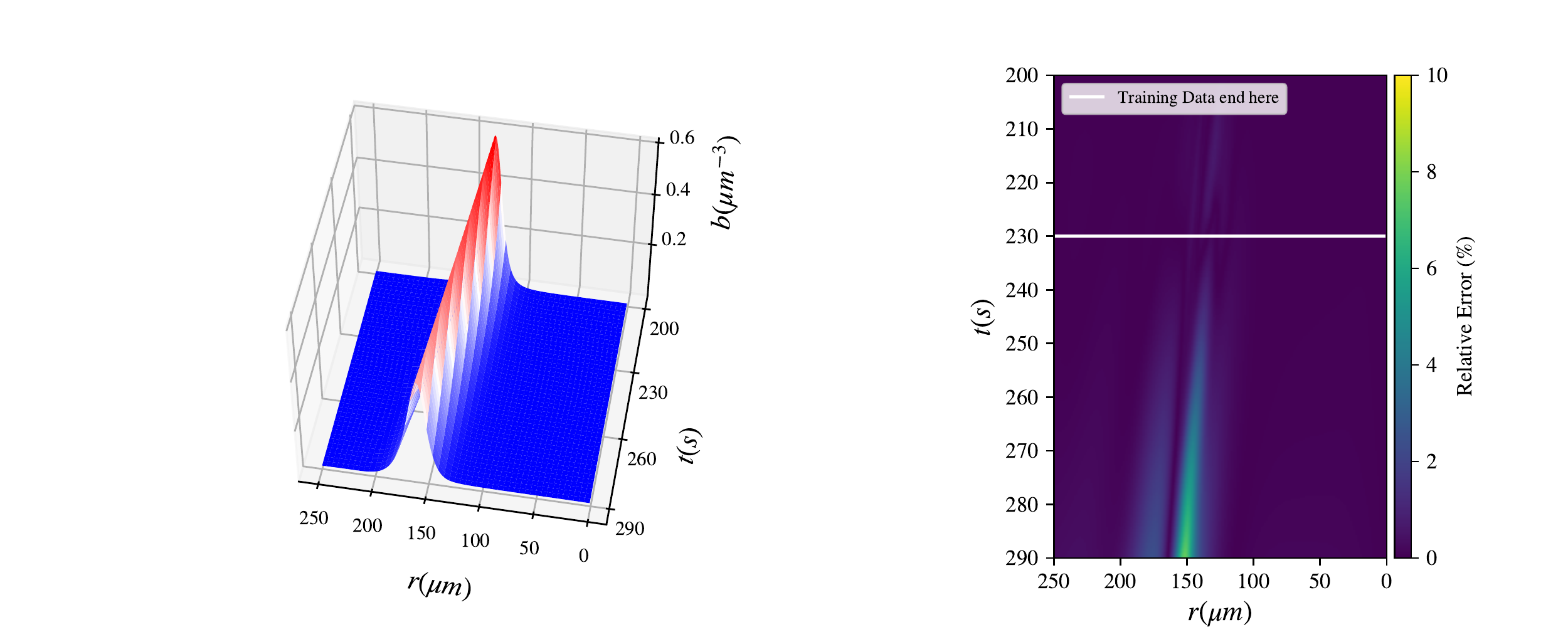}}
     \captionof{figure}{Black-box learning with a Neural Network: (left) Integration results for the data-driven PDE and (right)  \% relative error.}
     \label{fig:figNNBB}
\end{center}

Figure \ref{fig:figNNBB} showcases learning of \textit{one of the two} data-driven PDEs when the second PDE is known. Here, the chemonutrient ($c(r,t)$) PDE is assumed known, and the bacterial density evolution PDE is learned with a neural network, similar to
Eq.\ref{eq2NNBB}.

\vspace{1cm}


\paragraph{Black box ANN learning of a single field evolution equation, with only partial information (only the field $b(r,t)$ is observed)}   \hspace{2cm}\\  \label{SubSubSecBBNNpar}

After the initial success of the previous section, we decided to attempt a computational experiment, based on the spirit of the Takens embedding for finite-dimensional dynamical systems \cite{Takens_1981, Whitney_1936,  STARK19975303, Stark1999, Stark2003, Sauer1991}.

The idea here is that, if only observations of a few (or even only one) variables involved are available, one can use history of these observations (``time-delay" measurements) to create a useful latent space in which to embed the dynamics -and in which, therefore, to learn a surrogate model with less variables, but involving also history of these variables \cite{Packard1980, Aeyels1981}. 
There are important assumptions here: finite (even low) dimensionality of the long-term dynamics, something easier to contemplate for ODEs, but possible for PDEs with attracting, low dimensional, (possibly inertial) manifolds for their long-term dynamics. 
There is also the assumption that the variable whose values and history we measure is a {\em generic observer} for the dynamics on this manifold. 

One can always claim that, if a 100-point finite difference discretization of our problem is deemed accurate (so, for two fields, 200 degrees of freedom), then the current discretized observation of one of the two fields (100 measurements) plus three delayed observations of it ($3 \times 100$) plus possibly one more measurement give us enough variables for a useful latent space in which to learn a surrogate model. %
Here we attempted to do it with much less: at each discretization point we attempted keeping the current $b(r,t)$ field measurement and its spatial derivatives, and added only some history (the values and spatial derivatives at the previous moment in time). 
The equation below is written in discrete time form (notice the dependence of the field at the next time step from two previous time steps); it can be thought of as a discretization of a {\em second order in time} partial differential equation for the $b(r,t)$ field, based on its current value and its recent history. 

\begin{align}  \label{eqNNBBpar}
    b(t_{k+1})=b(t_k)+ \Delta t f^{partial}_{NN}( &b(t_k), (\mathbf{\nabla b} \cdot \mathbf{\hat{r}})(t_k),(\Delta b) (t_k), \nonumber \\ 
    & b(t_{k-1}),
    (\mathbf{\nabla b} \cdot \mathbf{\hat{r}})(t_{k-1}),(\Delta b)(t_{k-1}) ), 
\end{align}

\noindent with $\Delta t = t_{k+1} - t_k$, for any time point $t_k, k \geqslant 1$.


\begin{figure}
    \centering
    \makebox[0pt]{\includegraphics[width=20cm, height=8cm]{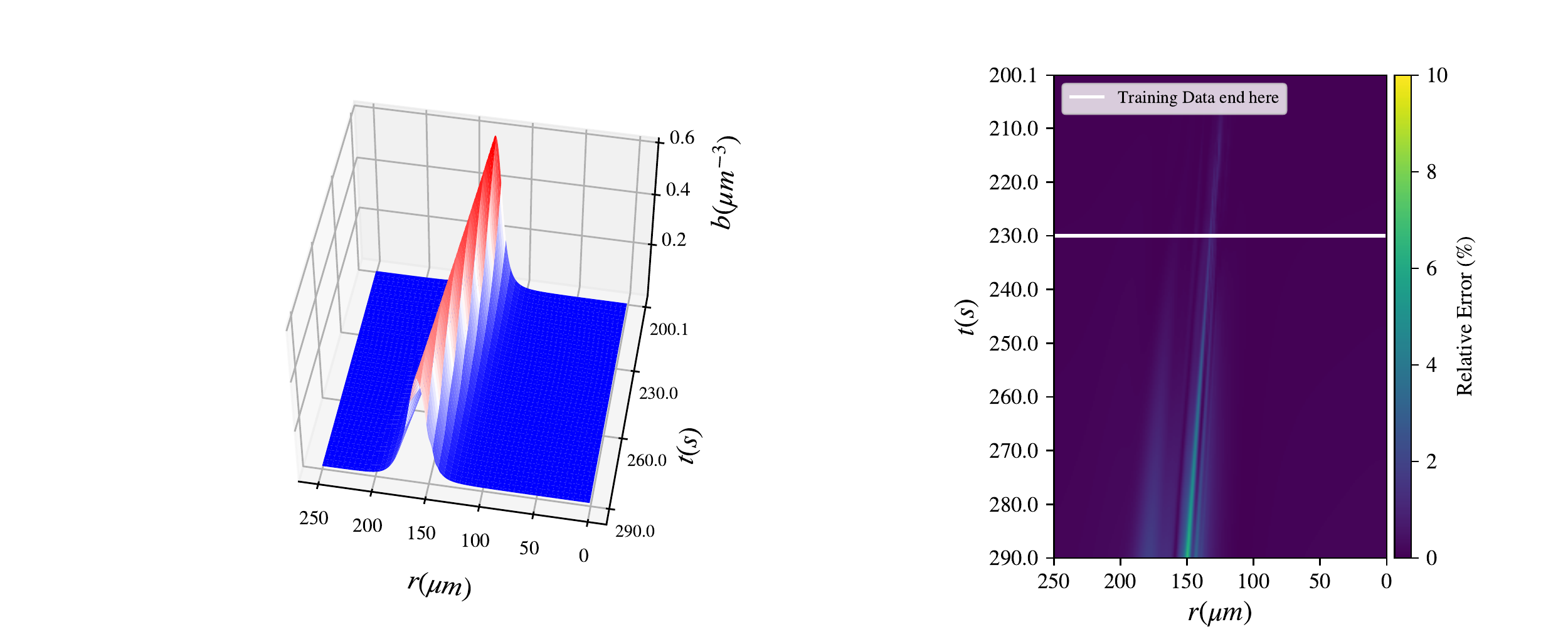}}
    \caption{Black-box partial-information learning with a Neural Network: (left) Integration results for the data-driven PDE and (right) \% relative error.}
    \label{fig:figNNBBpar}
\end{figure}

Figure \ref{fig:figNNBBpar} demonstrates learning a data-driven (discrete in time here) evolution equation for the bacterial density $b(r,t)$ when only data for $b(r,t)$ are at hand (partial information). Even though we knew the existence of another, physically coupled field, we assumed we cannot sample from it, so we replaced its effect on the $b(r,t)$ field through a functional dependence on the history of $b(r,t)$. Simulation of the resulting model was successful in reproducing the training data and extrapolating beyond them in time.

\subsubsection{Gray-box data-diven models}
\label{subsec:gbmodels}

A similar approach can be implemented when we have knowledge of a term/ some of the terms {\bf but not of the rest of the terms} of the right-hand side. In the specific context of chemotaxis, we are interested in learning just the chemotactic term, i.e. functions $g_i: \mathbb{R}^{3d\cdot m} \rightarrow \mathbb{R}^m, i \in \{1, ..., d\}$ such that:
\begin{equation} \label{ddPDE_GB}
    u^{(i)}_t - D^{(i)} \Delta u^{(i)} = g_i( u^{(1)}, ...,  u^{(d)},\mathbf{\nabla u^{(1)}} \cdot \mathbf{\hat{x}}, ..., \mathbf{\nabla u^{(d)}} \cdot \mathbf{\hat{x}},\Delta u^{(1)}, ...,  \Delta u^{(d)} ), \nonumber
\end{equation}

\noindent where $D^{(i)}$ is an \textit{a priori} known diffusivity. This is now a \textbf{gray box model} for the macroscopic PDE, and is particularly useful in cases where an (effective) diffusion coefficient is easy to determine, possibly from a separate set of experiments or simulations \cite{psarellis2022}. Again, as for black box models, gray boxes can also be formulated in the case of partial information, i.e. when not all fields $u^{(i)}$ are known, by leveraging history information of the known variables.

\vspace{1cm}


\paragraph{Gray box learning with ANN - $c_t$  known (with fields $b(r,t), c(r,t)$ known)}
\label{SubSubSecGBNN}

\begin{equation} \label{eqNNGB}
    b_t - D_b \Delta b = g_{NN}(b, \mathbf{\nabla b} \cdot \mathbf{\hat{r}},\Delta b, c, \mathbf{\nabla c} \cdot \mathbf{\hat{r}},\Delta c )
\end{equation}

\begin{center}
     \makebox[\textwidth][c]{\includegraphics[width=17.5cm, height=7cm]{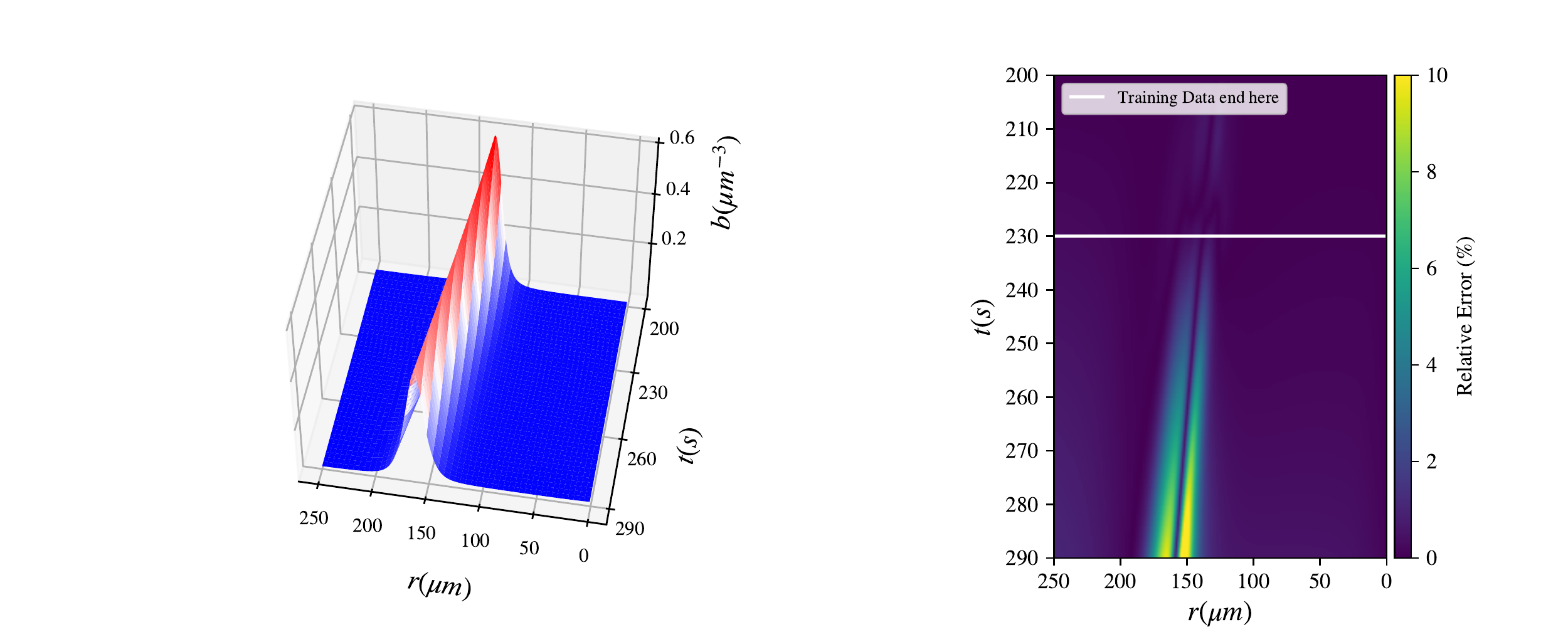}}
     \captionof{figure}{Gray-box learning with a Neural Network: (left) Integration results for the data-driven PDE and (right) \% relative error.}
     \label{fig:figNNGB}
\end{center}

Figure \ref{fig:figNNGB} shows the performance of gray-box models, where only the chemotactic term of the bacteria density PDE was considered unknown. For this gray-box model, the effective diffusion coefficient for the bacterial density was considered known. 
In principle, one could also hardwire the knowledge {\em of the functional form of this term} in the loss of the neural network, and thus obtain an estimate of this diffusivity in addition to learning the chemotactic term in the PDE.


\subsubsection{Estimating the chemonutrient field  - Computational Data.}
\label{SubsSecCNN}

Following up the above success in using Takens' embeddings (and more generally, Whitney embeddings) \cite{Takens_1981, Whitney_1936} for low-dimensional long-term dynamics, we attempted to estimate (i.e. create a nonlinear observer - a ``soft sensor" of) the chemonutrient field from local measurements of the bacterial fields and its history \cite{Altaf2017, Farhat2017}.
More specifically, we attempted to train a neural network to learn (in a data driven manner) $c(r_i,t_k)$
as a function of some local space time information:
\begin{align} \label{eqNNC_unnorm}
    c(r_i, t_k) = C_{NN}( & b(r_i, t_k), (\mathbf{\nabla b} \cdot \mathbf{\hat{r}})(r_i, t_{k}),  \nonumber
    (\Delta b) (r_i, t_k),\\ & b(r_i,t_{k-1}), (\mathbf{\nabla b} \cdot \mathbf{\hat{r}})(r_i, t_{k-1}),(\Delta b)(r_i, t_{k-1})), \tag{6a}
\end{align}

\noindent for any discretization point is space $r_i$ and time point $t_k, k \geqslant 1$. \\

Indeed, if the long-term simulation data live on a low-dimensional, say $m-$dimensional manifold, then $2m+1$ generic observables suffice to embed the manifold, and then learn {\em any} function on the manifold in terms of these observables.
Here we attempted an experiment with a neural network that uses
a {\em local} parametrization of this manifold, testing if such a local parametrization can be learned (in the spirit of the nonlinear discretization schemes of Brenner et al \cite{Brenner2019}).

There is, however, one particular technical difficulty: because the long-term dynamics of our problem appear in the form of travelling waves, both the front and the back of the wave appear practically {\em flat} -- the spatial derivatives are all approximately zero, and a simple neural network cannot easily distinguish, from local data, if we find ourselves in the flat part {\em in front} of the wave or {\em behind} the wave. 
We therefore constructed an additional local observable, capable of discriminating between flat profiles ``before" and flat profiles ``after" the wave. 
Indeed, when the data represent the spatiotemporal evolution of a traveling wave (as in our training/testing data set), we expect a \textbf{singularity close to} $\mathbf{b(r,t)=0}$. 
Clearly, however, the $c$ field is dramatically different on the two ``flat bacteria" sides (see the left panel of Fig.\ref{fig:transformation}). When learning such a function \textbf{locally}, to circumvent this singularity, we proposed using a  transformation of two of the inputs:
$(b, \mathbf{\nabla b} \cdot \mathbf{\hat{r}}) \rightarrow \left(b, \arctan(\frac{\overline{(\mathbf{\nabla b} \cdot  \mathbf{\hat{r}})}} {\overline{b}} \right) $, where the bar symbol denotes an affine transformation of the respective entire feature vector to the interval $[-1,1]$. This transformation brings points at different sides of the aforementioned singularity at different ends of a line, exploiting their difference in sign (see Fig.\ref{fig:transformation}).

\begin{center}
     \makebox[\textwidth][c]{\includegraphics[width=18cm, height=6cm]{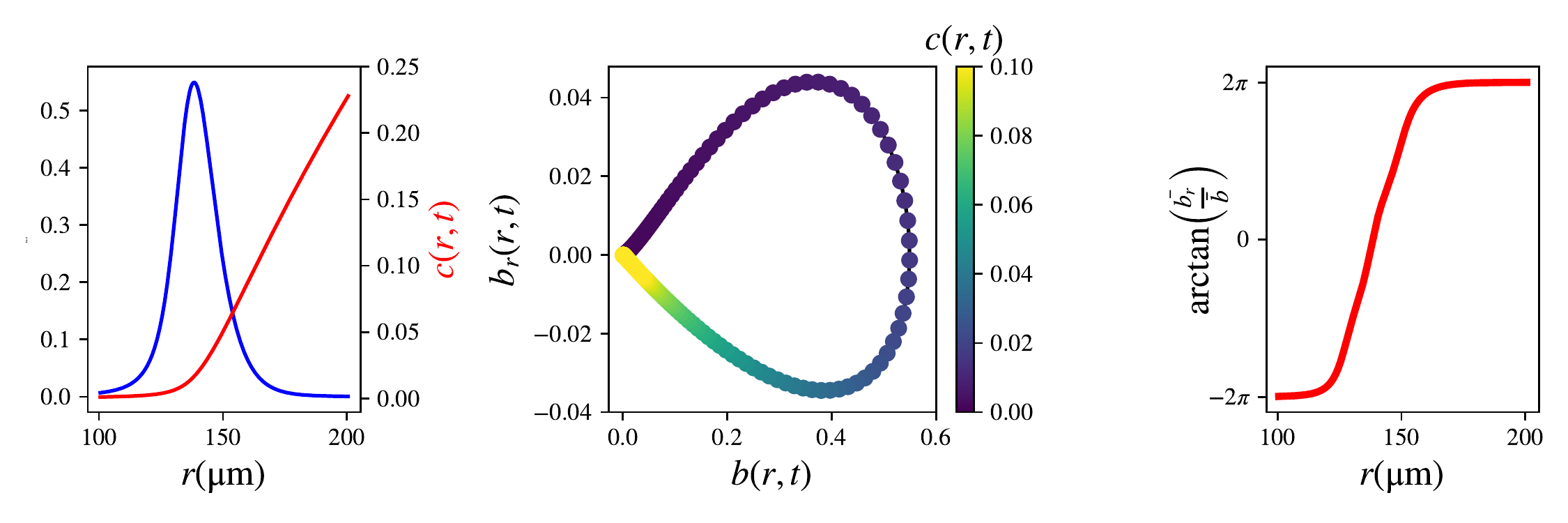}}
     \captionof{figure}{Transformation on the inputs: (left) representative profiles of both fields $b(r,t), c(r,t)$, (middle) visualization of the singularity, (right) transformed variable.}
     \label{fig:transformation}
\end{center}



Then, the Neural Network was trained to learn the estimator (nonlinear observer) of the chemonutrient field as:

\begin{align} \label{eqNNC}
    c(r_i, t_k) = C_{NN}\Bigl(&b(r_i, t_k), \arctan\left(\frac{\overline{(\mathbf{\nabla b} \cdot    \nonumber
    \mathbf{\hat{r}})(r_i, t_k)}} {\overline{b(r_i, t_k)}}\right),(\Delta b) (r_i,t_k),\\  \nonumber
    & b(r_i, t_{k-1}), (\mathbf{\nabla b} \cdot \mathbf{\hat{r}})(r_i,t_{k-1}),(\Delta b)(r_i,t_{k-1})\Bigr), \tag{6b}
\end{align}

for any discretization point is space $r_i$ and time point $t_k, k \geqslant 1$.

\begin{center}
     \makebox[\textwidth][c]{\includegraphics[width=20cm, height=8cm]{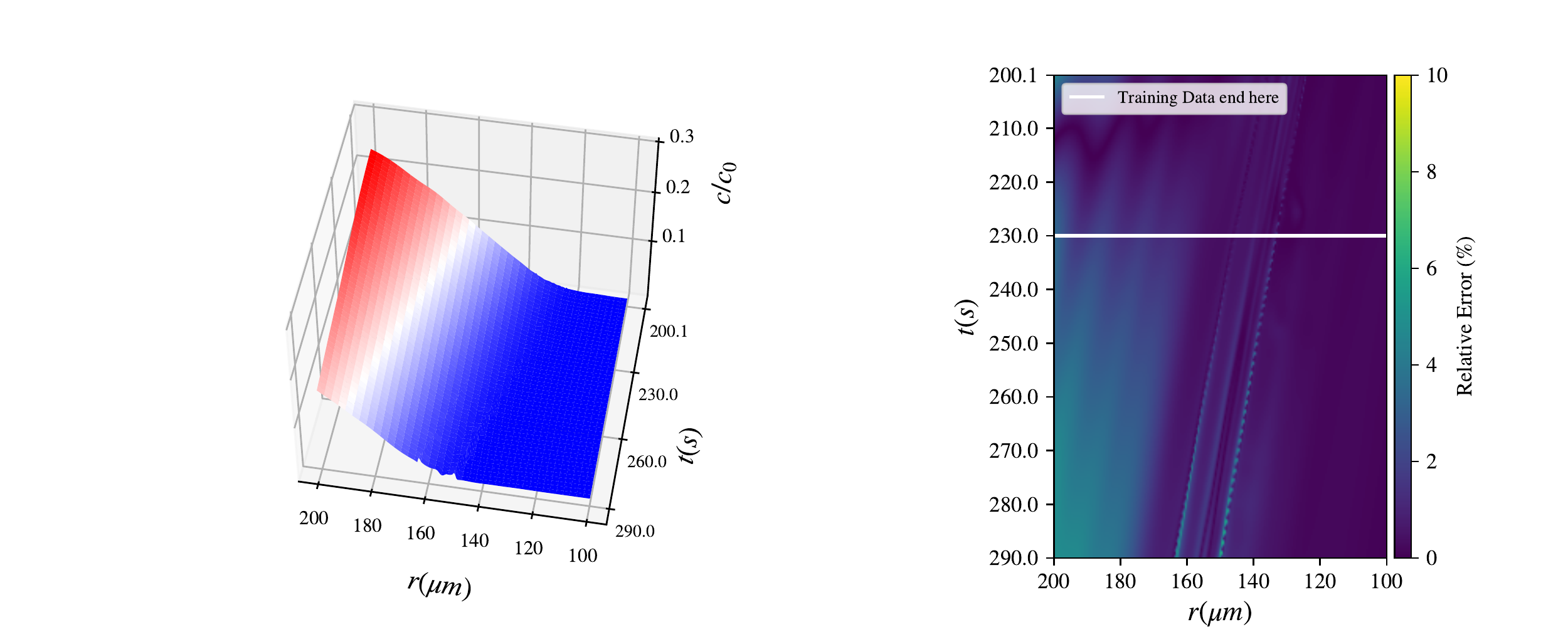}}
     \captionof{figure}{Learning the c-field with a Neural Network: (left) Field prediction and (right) \% relative error.}
     \label{fig:figNNc}
\end{center}

As it can be seen from Fig.\ref{fig:figNNc} through the model in Eq.\ref{eqNNC} it was possible to provide reasonable predictions for the chemonutrient field.


\subsection{Black box model - Experimental data}
\label{SecResultsExp}

The chemotactic motion of bacteria can also be studied through laboratory experiments. As shown in \cite{Bhattacharjee_2020}, chemotactic motion can be tracked using confocal fluorescence microscopy of \textit{E. coli} populations; thus, we used the data from these prior experiments here. As detailed further in \cite{Bhattacharjee_2020}, we 3D-printed a long, cylindrical inoculum of densely-packed cells within a transparent, biocompatible, 3D porous medium comprised of a packing of hydrogel particles swollen in a defined rich liquid medium. Because the individual particles were highly swollen, their internal mesh size was large enough to permit free diffusion of oxygen and nutrient (primarily \textit{L}-serine, which also acts as a chemoattractant), but small enough to prevent bacteria from penetrating into the individual particles; instead, the cells swam through the interstitial pores between hydrogel particles.

We hypothesized that the spatiotemporal behavior of cell density observed in the experiments results from a PDE similar to the one used in simulations (Eq. \ref{eq:1}). However, having no measurements of the spatiotemporal evolution of the chemonutrient, we turned to the methodology described earlier for data-driven models with partial information (Sec. \ref{SubSubSecBBNNpar}). 


Interpolation in time allowed for well-approximated time derivatives as we could choose data along $t$ as dense as necessary. In fact, it was possible to estimate second order in time derivatives, which could be used to learn a second order in time continuous-time PDE in lieu of a delay model \cite{Packard1980}, such as that used in \ref{eqNNBBpar}:

\begin{equation} \label{eqNNexp}
    b_{tt} = f^{exp}_{NN}(b, \mathbf{\nabla b} \cdot \mathbf{\hat{r}},\Delta b, b_t) \tag{7}
\end{equation}

This was treated as two coupled first-order PDEs, using the intermediate ``velocity'' field $u(r,t)$:

\begin{align} 
    u_{t} &= f^{exp}_{NN}(b, \mathbf{\nabla b} \cdot \mathbf{\hat{r}},\Delta b, u) \nonumber \\
    b_t &= u \label{eqNNexpint}
\end{align}

Given the nutrient-starved/hypoxic conditions at $r \approx 0$ (see Section \ref{methods_exp_desc}), our training data were selected away from the origin. We prescribed bilateral boundary corridors to provide data-informed boundary conditions when integrating the learned PDE.

\vspace{2cm}

\begin{center}
    \makebox[\textwidth][c]{\includegraphics[width=9cm, height=8cm]{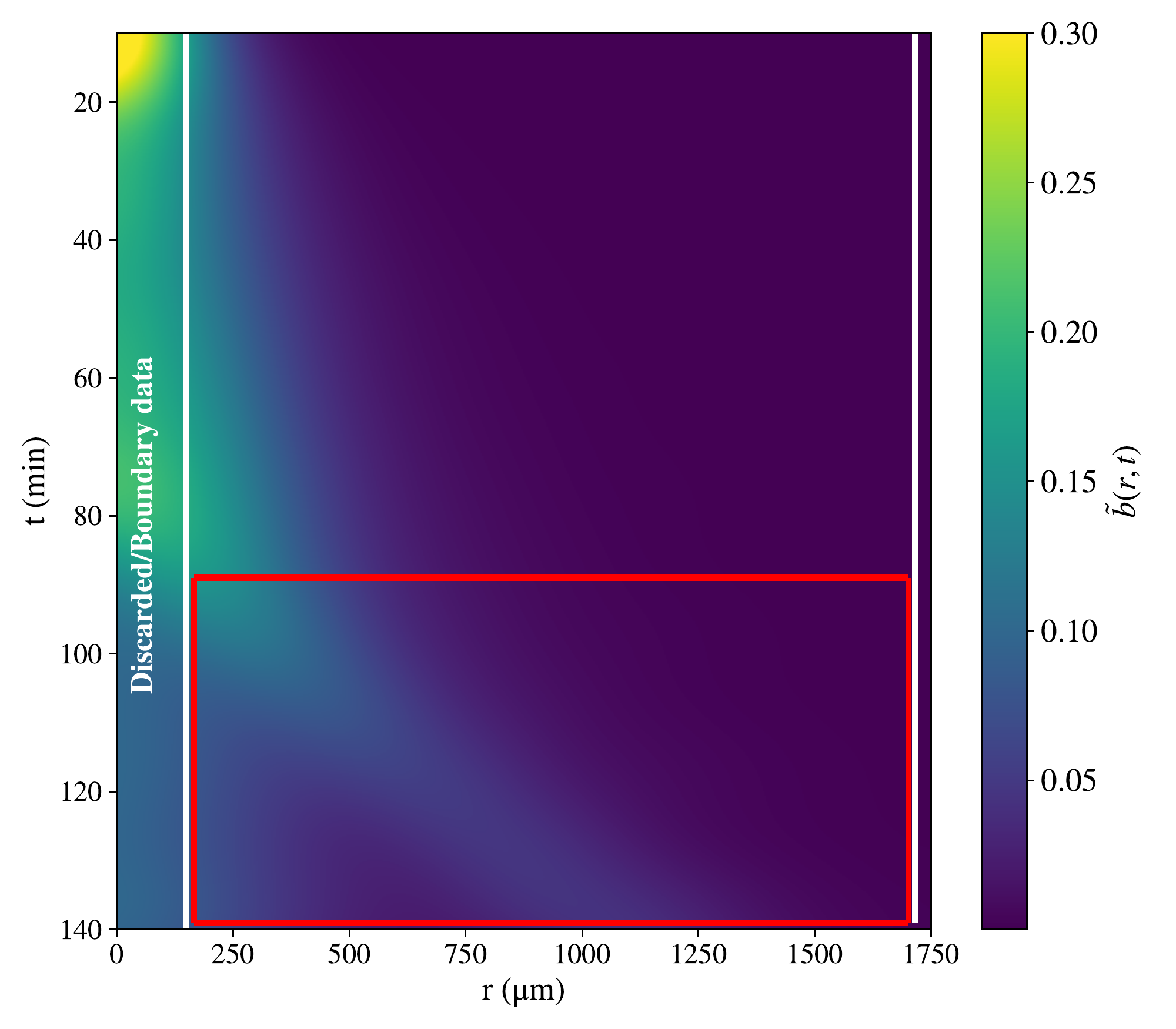}}
    \captionof{figure}{Segmentation of the pre-processed data into: boundary corridors/ discarded data, training subset (the complement), subset chosen for reproduction (red rectangle).} 
\label{fig:boundary_cor}
\end{center}

\begin{center}
    \makebox[\textwidth][c]{\includegraphics[width=22cm, height=6.6cm]{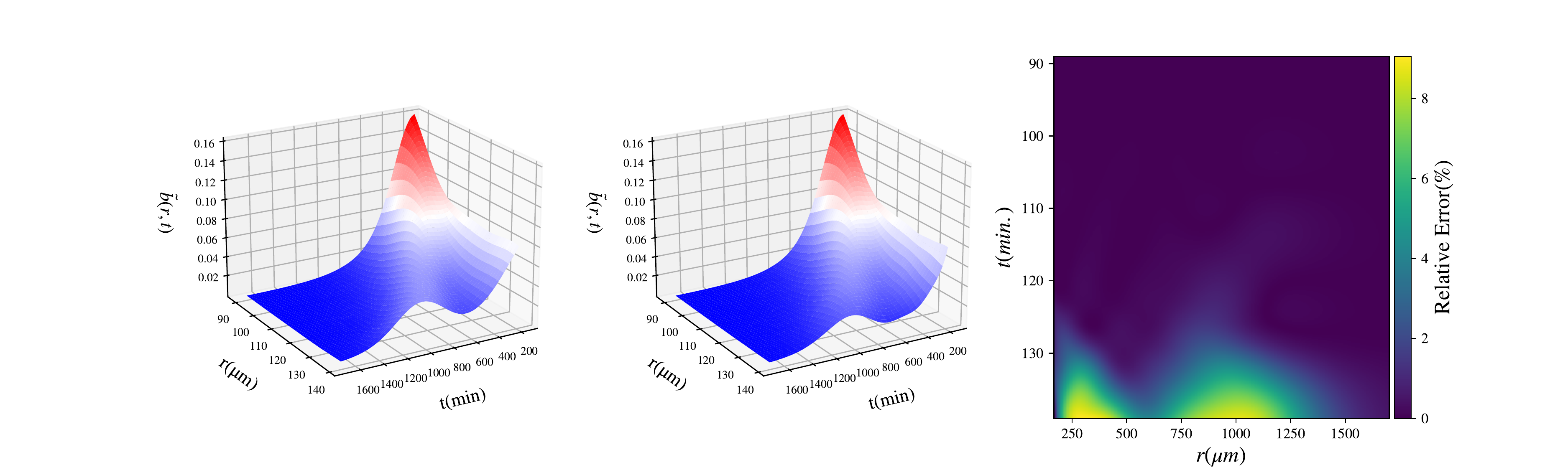}}
    \captionof{figure}{Reproduction of experimental observations using a Data-driven Neural Network for the traveling wave regime: (left) Ground Truth, (middle) Neural-Network PDE integration results, (right) \% Relative Error.} 
\label{fig:figres}
\end{center}

The model was validated by integration in the spatiotemporal domain of the formation and propagation of the traveling wave (shown in red in Fig. \ref{fig:boundary_cor}). Integration results can be seen in Fig. \ref{fig:figres}.

\section{Discussion}
\label{SecDisc}

We demonstrated here how data-driven methods are able to learn macroscopic chemotactic PDEs from bacterial migration data. The methodology presented was applied either to data sets from fine PDE simulations or from experiments. For an example with agent-based computations see \cite{psarellis2022}. The same task can be accomplished even when the data at hand are partial, noisy, and/or sparse. It is also possible to learn just one term of a PDE, or a certain PDE out of a set of coupled PDEs.  These data-driven models were able to reproduce spatiotemporal profiles used for training and extrapolate further in time. This work showcases that data-driven PDEs are a versatile tool which can be adjusted and implemented to many different problem settings, data sets or learning objectives. It can be especially useful (if not necessary) when the derivation of an analytical PDE is cumbersome, or when there is no capacity for a large number of simulations or experiments. In fact, data-driven PDEs can be used to estimate transients from different initial conditions (IC), for different boundary conditions (BC) or different spatial domain geometries. Apart from that, learning data-driven PDEs is one of the most compact and generalizable ways to learn a system's behavior from data. A lot of more specific observations can be drawn from comparing results of different models or methodologies in Section \ref{SecMethods_results}.

\begin{itemize}
    \item For the model in Eq.\ref{eq2NNBB}, where both PDEs are data-driven it is obvious that the $c(r,t)$ PDE is easier to learn and predict than the $b(r,t)$ PDE, both for the ANN and the GPR (see SI). This can be understood in terms of complexity of these two target functions: The $b_t$ expression is highly nonlinear (owing mostly to the logarithmic chemotactic term) and complex, as it depends on most of the input features. On the contrary the $c_t$ expression only depends on a handful of inputs and is less complicated.
    \item When learning both PDE right-hand sides (model in Eq.\ref{eq2NNBB}), it is
    straightforward to employ an ANN for multiple output prediction. However, multiple output GP (or cokriging) is especially non-trivial \cite{WANG2015159} and therefore, multiple single-output GPs are preferred instead.
    \item When comparing the ANN methodology with the GPR (in SI), for the same model, it can be seen that ANNs tend to produce more accurate predictions than GPR. This may be attributed to the ANN's versatility and efficiency in capturing the nonlinearities and complexities of any target function. It is also important to note that, due to memory constraints, GPR was trained only with an appropriately chosen subset of the training data. This could cause the loss of accuracy in long term predictions. However, it is important to note, that the error in GPR is always \textit{smooth}, owing to the smoothness of the Gaussian kernel (see SI).
    \item Comparing the predictions for $b_t$ from models in Eqs.\ref{eq2NNBB} and \ref{eqNNBB} it can be seen that when only \textit{one} of the PDEs is data-driven, the predictions are more accurate. This may be rationalized in terms of error accumulation during integration: when both PDEs are data-driven, both of the PDEs contribute to prediction error which will propagate along the integration trajectory.
    \item  The models with partial information were trained with a single (therefore, fixed) delay ($t_k-t_{k-1}$), which imposes important restrictions on how we can advance in time. A natural way to do this, is with a Forward Euler scheme with a timestep equal to the delay used in training, as explicitly shown in Eq.\ref{eqNNBBpar}.

\end{itemize}

Observations for models for simulations data (Section \ref{SecResultsExp}) are aligned with those for the experimental data (Section \ref{subsec:sim}): A second order model can indeed capture the dynamics of a data set with partial information. In this case, a deeper Neural Network is required to capture the real-world dynamics from experimental observations. The data-driven PDE manages to capture important characteristics of the traveling wave, such as its speed and amplitude. It also manages to capture the dynamics \textit{on the left} of the traveling wave: the bacteria density remains stationary, as in that region the chemonutrient gradient is negligible. Note that, as discussed in \cite{Bhattacharjee_2020}, analytical models fail to capture this behavior without non-autonomous correction terms. 

Future work includes relaxing some of the assumptions used here; for example, it is possible to train these models in a \textit{coordinate - free} way \cite{luk2020a, weiler2021coordinate}. This would result in data-driven PDEs which are valid under coordinate changes (e.g. Cartesian, polar, spherical). That is possible by expressing the PDE in terms of the exterior derivatives. Please refer to the SI for a brief outlook on the use of differential operators arising from exterior calculus~\cite{flanders1989,lee2009} to create a dictionary of features in which to express learned operators.

In addition, it is also possible to limit the number of independent, relevant inputs in a data-driven way, using dimensionality reduction, automatic relevance determination or other feature importance methods \cite{rasmussen:williams:2006, Ghorbani_Abid_Zou_2019}. These future directions aim in more robust and generalizable data-driven PDEs.

\section{Materials and Methods}
\label{SecChemo}

\subsection{PDE simulations}
\label{methods_pde}

\begin{figure}[h]
    \begin{tabular}{||c c c||} 
 \hline
 Parameter &  Value & Unit \\ [0.5ex] 
 \hline\hline
 $D_b$ & $2.325$ & $\mathrm{\upmu m^2 /s}$ \\
 $\chi_0$  & $17.9$ & $\mathrm{\upmu m^2 /s}$ \\
 $c_-$  & $1$ &  $\mathrm{\upmu M}$ \\
 $c_+$  & $30$ & $\mathrm{\upmu M}$ \\
 $c_{char}$ & $1$ & $\mathrm{\upmu M}$ \\
 $\gamma$  & $0$ & $\mathrm{\upmu M/s/ \upmu m^3}$ \\
 $D_c$  & $800$ &  $\mathrm{\upmu m^2 /s}$ \\
 $\kappa$  & $3000$ &  $\mathrm{\upmu M/s/ \upmu m^3}$ \\
 $b_0$  & 0.95 &  $ 1/ \mathrm{\upmu m^3}$ \\
 $\sigma$  & 42.62 &  $\mathrm{\upmu M}$ \\
 $c_0$  & $10$ &  $\mathrm{m M}$ \\
 $R$  & $17.5$ &  $\mathrm{m m}$ \\[1ex] 
 \hline
\end{tabular}
    \includegraphics[width=5cm, height=4cm]{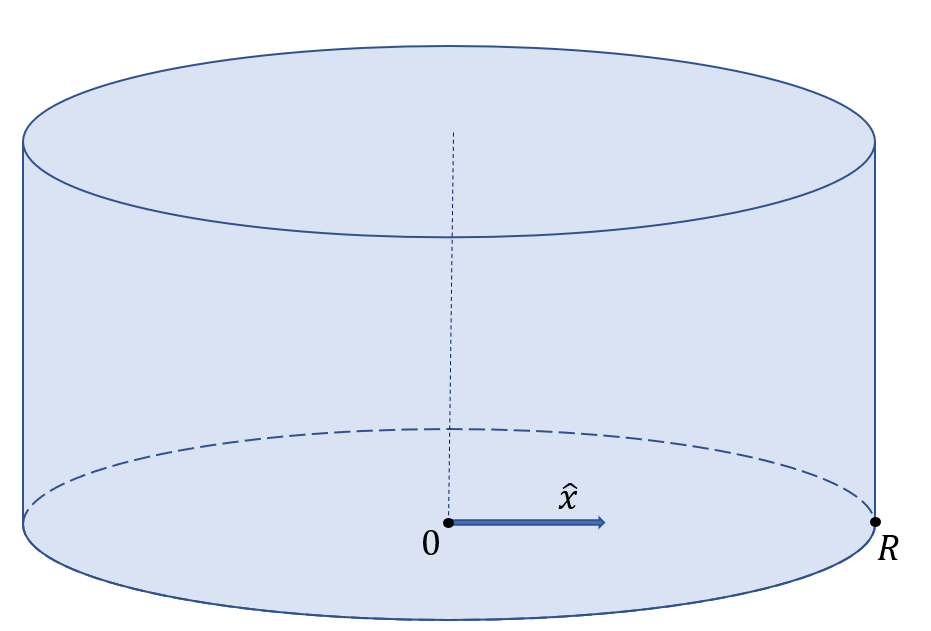}
    \label{fig:sch}

    \captionlistentry[table]{Parameters used for the PDE simulation and a schematic representation of the domain.}     \label{table:2}

    \captionsetup{labelformat=andtable}
    \caption{Parameters used for the PDE simulation of the extended Keller-Segel model and a schematic representation of the computational domain.}
 \label{figtab}
 \end{figure}

To model chemotactic motion of \textit{E. coli} in heterogeneous porous media, the following extension of the Keller-Segel model \cite{Bhattacharjee_2020} was used:

\begin{align} 
    & \frac{\partial{b}}{\partial{t}} = D_b \Delta b -\chi_0\nabla\cdot [b \nabla log F_1(c)] +b \gamma F_2(c)  \nonumber \\
    & \frac{\partial{c}}{\partial{t}} = D_c \Delta c -b \kappa F_2(c)  \nonumber \\
    & \mathbf{\nabla J_b}(0,t)\cdot \hat{\mathbf{n}} = 0,  \mathbf{\nabla J_b}(R,t)\cdot \hat{\mathbf{n}} = 0  \nonumber \\
    & \mathbf{\nabla c}(0,t) \cdot \hat{\mathbf{n}} = 0,  \mathbf{\nabla c}(R,t)\cdot \hat{\mathbf{n}} = 0,  \label{eq:1}
\end{align}

\noindent  where in radial coordinates $b(r,t)$ is the bacterial density, $c(r,t)$ is the chemonutrient concentration, $D_b$ is the bacterial diffusion coefficient, $D_c$ is the chemonutrient diffusion coefficient, $R$ (in the boundary conditions) is the overall domain radius, $F_1(c)=\frac{1+c/c_-}{1+c/c_+}$ and $F_2(c)= \frac{c}{c+c_{char}}$,  $\mathbf{J_b}$ is the bacterial density flux ($\mathbf{J_b} = D_b \nabla b - \chi_0 b \nabla log F_1(c)) $   and $\hat{\mathbf{n}}$ is the normal vector at the domain boundaries. Note that in this case, no flux boundary conditions imply zero gradients for both fields at the circular/cylindrical boundary.
Experiments can be performed to individually estimate the parameters $D_b, \chi_0, c_-, c_+, D_c, c_{char}, \gamma$, as in \cite{Bhattacharjee_2020} (also see Table \ref{table:2}).

Numerical simulations of Eq.\ref{eq:1}, to provide training data for our data-driven
identification approach, were performed using COMSOL Multiphysics 5.5 \cite{multiphysics1998introduction}, for the spatiotemporal domain $(t,r) \in [0, 300] \times [0, 1000]$, with initial conditions $b(r, 0) = b_0 e^{-r^2 /2\sigma^2}, c(r, 0) =c_0$. 
Note  that for all learning cases, the training data are (a suitable subset) in  $(t,r)  \in  [200, 230] \times [0, 250]$ (therefore, $m\cdot n \leq 15,801$ in $\mathbb{R}^2$) and the model is validated by simulation in  $(t,r) \in [200, 290] \times [0, 250]$. with spatial resolution $dr= 0.5 \mu m$ with the results reported every $\delta t =0.1s$ (relative tolerance set at $10^{-7}$). Integration was performed using Finite Element Method and a MUMPS solver \cite{multiphysics1998introduction}.The parameters of Eq.\ref{eq:1} used in the simulation can be found in Table  \ref{table:2}.

\subsection{Description of the experiments}
\label{methods_exp_desc}

The cells constitutively express fluorescent protein in their cytoplasms, enabling us to track their motion as they expanded radially outward from the initial cylindrical inoculum in 3D. The fluorescence measurements were collected with spatial resolution $dr =2.48~\mu m$ and temporal resolution $dt =10~min$, and were then azimuthally averaged, only considering signal from the transverse, not the vertical, direction. The fluorescence signal thereby determined is directly proportional to the density of metabolically-active bacteria, and is denoted as $\Tilde{b}(r,t)$; cells are left behind in the wake of the moving chemotactic front, but become immobilized and lose fluorescence as they run out of oxygen and nutrients. More information can be found in \cite{Bhattacharjee_2020}.

\subsubsection{NN Training for a sample model}
\label{methods:NN}

As an example, we mention here the training details regarding one of the models, the one in Eq.\ref{eq2NNBB}. Training was performed with a feedforward neural network consisting of two hidden layers, each with 18 tanh-activated neurons. An Adams optimizer \cite{adams2014} was used with a MSE loss function. The neural network hyperparameters were empirically tuned: 2048 epochs and 0.02 learning rate. After training, the data-driven PDE was integrated with a commercial BDF integrator, as implemented in Python's {\fontfamily{qcr}\selectfont
solve\_ivp} \cite{BDF1997}, with relative and absolute tolerances at $(10^{-4}, 10^{-7})$. The initial profile for integration was supplied by our simulation data, and the boundary conditions set to no flux. The Jacobian of the data-driven PDE was provided via automatic differentiation.

\subsubsection{Preprocessing and NN Training for experimental data}
\label{methods:exo_train}

The training profiles were selected appropriately so that the traveling wave is not too close to the spatial boundaries, and the cylindrical coordinate system remains valid. Profiles were smoothed in space using a local Savitzky-Golay filter and globally using Gaussian Smoothing \cite{savitzky, Gaussian}. The resulting smooth profiles were interpolated in time using Gaussian Radial Basis Functions \cite{rbf}.

\begin{center}
    \makebox[\textwidth][c]{\includegraphics[width=20cm, height=8cm]{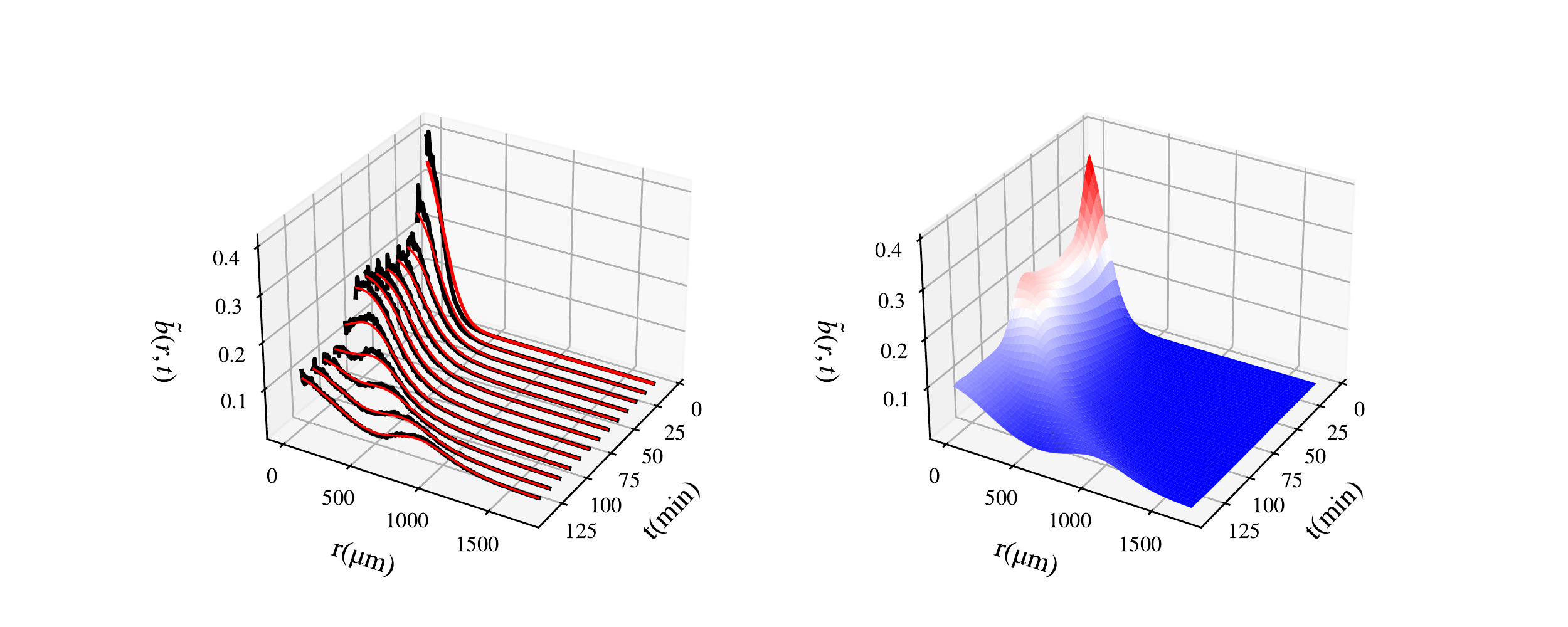}}
    \captionof{figure}{Pre-processing of experimental measurements: (left) Smoothing in space (right) Interpolation in time.}
\label{fig:figsmooth}
\end{center}

The learning algorithm consists of a Deep feed-forward Neural Network with 3 hidden layers and 90 neurons per layer. When integrating the data-driven model, SVD filtering was used: $u_t$ is projected to a lower-dimensional space, defined by the dominant singular vectors of the the $u_t(r,t)$ data used in training \cite{halko2010finding}.  This is a procedure analogous to adding hyperviscosity
in hydrodynamic models \cite{Thiem_2021}. Here, the eight first singular vectors were used, containing $>99\%$ of the variance.
\clearpage
\bibliography{Bibliography}

\end{document}


\title{Supporting Information: \\
Data-driven Discovery of Chemotactic Migration of Bacteria via Machine Learning}

\author[jhu]{Yorgos M. Psarellis}
\author[sjsu]{Seungjoon Lee}
\author[princeton2]{Tapomoy Bhattacharjee}
\author[princeton]{\newline Sujit S. Datta}
\author[jhu]{Juan M. Bello-Rivas}
\author[jhu,jhu1,jhu2]{Ioannis G. Kevrekidis\corref{mycorrespondingauthor}}
\cortext[mycorrespondingauthor]{Corresponding author}
\ead{yannisk@jhu.edu}

\address[jhu]{Department of Chemical and Biomolecular Engineering, Johns Hopkins University}
\address[sjsu]{Department of Applied Data Science, San Jos{\'e} University}
\address[princeton2]{Andlinger Center for Energy and the Environment, Princeton University}
\address[princeton]{Department of Chemical and Biological Engineering, Princeton University}
\address[jhu1]{Department of Applied Mathematics and Statistics, Johns Hopkins University}
\address[jhu2]{Department of Medicine, Johns Hopkins University}
\maketitle

\section{Gaussian Process Regression}
\label{SecGP}

To learn a function $f$ from data we can employ Gaussian Process Regression (GPR). GPR assumes that the target function $f(\mathbf{x}), f:\mathbb{R}^n \rightarrow \mathbb{R}$ is distributed according to a Gaussian process, which can be fully specified by its mean function $m(\mathbf{x})$ and covariance function $k(\mathbf{x}, \mathbf{x'})$ \cite{rasmussen:williams:2006}:
\begin{equation} \label{gp}
    f(\mathbf{x}) \sim \mathcal{N}(m(\mathbf{x}), k(\mathbf{x}, \mathbf{x'})). \nonumber
\end{equation}


This can be understood as setting a Gaussian prior distribution over the space of functions. The mean is usually set to zero by centering the data during preprocessing. The covariance function $k(\mathbf{x}, \mathbf{x'})$ is commonly formulated by a Euclidean-distance kernel function in the input space \cite{Lee_2019}. Here, we use the Matèrn32 kernel with a constant:

\begin{align} \label{matern}
    k(\mathbf{x_i}, \mathbf{x_j}) &= c\left(1+ \sqrt{3} d(\mathbf{x_i}, \mathbf{x_j};\mathbf{l} )\right) e^{-\left(\sqrt{3}d(\mathbf{x_i}, \mathbf{x_j};\mathbf{l})\right)},  \\
    d(\mathbf{x_i}, \mathbf{x_j};\mathbf{l}) &= \sqrt{ \sum_{k=1} ^n \frac{({x_i}_k -{x_j}_k)^2}{l_k }}
\end{align}

\noindent where $\mathbf{x}_i, \mathbf{x}_j$ are any two feature vectors, $c$ is a scalar, $\mathbf{l}$ is a vector with number of entries equal to the dimension of the input space. $c$ and $\mathbf{l}$ are the hyperparameters to be optimized (here, collectively denoted $\boldsymbol {\theta}$).

Here, we consider the case of noisy observations: $y =f(\mathbf{x}) + \epsilon$, where $\epsilon \sim \mathcal{N}(0, \sigma^2)$ is i.i.d. Gaussian additive noise with known variance. Given a  dataset $\{(\mathbf{x}_i, y_i) | i = 1, ..., n\}$ the optimal hyperparameter vector $\bold{\theta^*}$ is the maximum likelihood estimator:
\begin{equation} \label{argmin}
    \mathbf{\theta^*}= \arg \min_{\mathbf{\theta}}{\{-log p(\bold{y} | \bold{x}, \mathbf{\theta)}}\} \nonumber
\end{equation}

This estimator defines the posterior Gaussian Process given $( \bold{x},\bold{y})$.
To find the Gaussian distribution of the function values at test data points, we represent the multivariate Gaussian distribution with the covariance matrix constructed by equation \ref{matern} as:

\begin{equation} \label{gp_predic1}
    \left[ \begin{array}{l}
    \bold{y} \\
    \bold{y^*}  \end{array} \right]  \sim N \left(\bold{0}, \\
    \left[ \begin{array}{cc}   \mathbf{K}+ \sigma^2 \mathbf{I} & \mathbf{K}_*  \\   \mathbf{K}^T_* & \mathbf{K}_{**}  \end{array}  \right]\right), \nonumber
\end{equation} 
\vspace{0.5cm}

\noindent where $\bold{y}^*$ is a predictive distribution for test data $\bold{x}^*$,  $\mathbf{K}_*$ represents a covariance matrix between training and test data while $\mathbf{K}_{**}$ represents a covariance matrix between test data.
Finally, we represent a Gaussian distribution for the target function at the test points in terms of a predictive mean and its variance, through conditioning a multivariate Gaussian distribution as:

\begin{align} \label{gp_predic2}
    & \bar{\bold{y}}^*= \mathbf{K}_*(\mathbf{K}+\sigma^2 \mathbf{I})^{-1} \bold{y}  \\
    & \mathbf{K}(\bold{y}^*) = \mathbf{K}_{**}- \mathbf{K}^T_*(\mathbf{K}+\sigma^2 \mathbf{I})^{-1} \mathbf{K}_*, \nonumber
\end{align} 

\noindent and we assign the predictive mean ($\bar{\bold{y}}^*$) as the estimated target function for the corresponding data point.

\section{Artificial Neural Networks}
\label{SecANN}


Artificial Neural Networks (ANNs) are a family of functions constructed by composing many affine and nonlinear elementary functions (activation functions). In (feed-forward) neural networks, a universal approximation theorem \cite{Cybenko89} guarantees that for a single hidden layer with (sufficient) finite number of neurons, an approximation $\Tilde{y}$ of the target function, $y$ can be found. Here, approximation implies that the target and learned functions are sufficiently close in an appropriately chosen norm: for all $ \epsilon >0$ there exists an ANN predicting $\Tilde{y}(\mathbf{x})$, where $ : |y(\mathbf{x}) -\Tilde{y}(\mathbf{x})| <\epsilon$  for all $\mathbf{x} \in A$ and some $A \subseteq \mathbb{R}^d$. The approximate form of the target function obtained through the feedforward neural net can be written as:

\begin{equation} \label{NN}
    \Tilde{y}(\mathbf{x}) =\sum_{i=1} ^{N_n} \psi(\mathbf{w}_i^T \mathbf{x} + \mathbf{b}_i), \nonumber
\end{equation}

\noindent where $\psi$ is a nonlinear activation function, $\mathbf{w}_i , \mathbf{b}_i$ are tunable parameters (weights and biases) and $N_n$ is the number of neurons, which is decided a priori. To find optimal weights and biases, an optimizer is used (employing a backpropagation scheme) to minimize the root-mean-square error cost function:

\begin{equation} \label{MSE}
    E_D=\frac{1}{n} \sum_{j=1}^n (y_j -\Tilde{y}(x_j))^2, \nonumber
\end{equation}
which typically measures the goodness of the approximation.

\section{PDE models not descibed in the text}
\label{SecRest}






\subsection{Black box learning of both PDEs with GPR (with fields $b(r,t), c(r,t)$ known).}
\label{SubSubSec2BBGP}

 \begin{align} \label{eq2GPBB}
   \begin{bmatrix}
           b_t \\
           c_t \\
           \end{bmatrix} &= \begin{bmatrix}
           f_{GP}(b, \mathbf{\nabla b} \cdot \mathbf{\hat{r}},\Delta b, c, \mathbf{\nabla c} \cdot \mathbf{\hat{r}},\Delta c ) \\
           h_{GP}(b, \mathbf{\nabla b} \cdot \mathbf{\hat{r}},\Delta b, c, \mathbf{\nabla c} \cdot \mathbf{\hat{r}},\Delta c ) \\
           \end{bmatrix} 
  \end{align}

\begin{center} 
     \makebox[\textwidth][c]{\includegraphics[width=17.5cm, height=7cm]{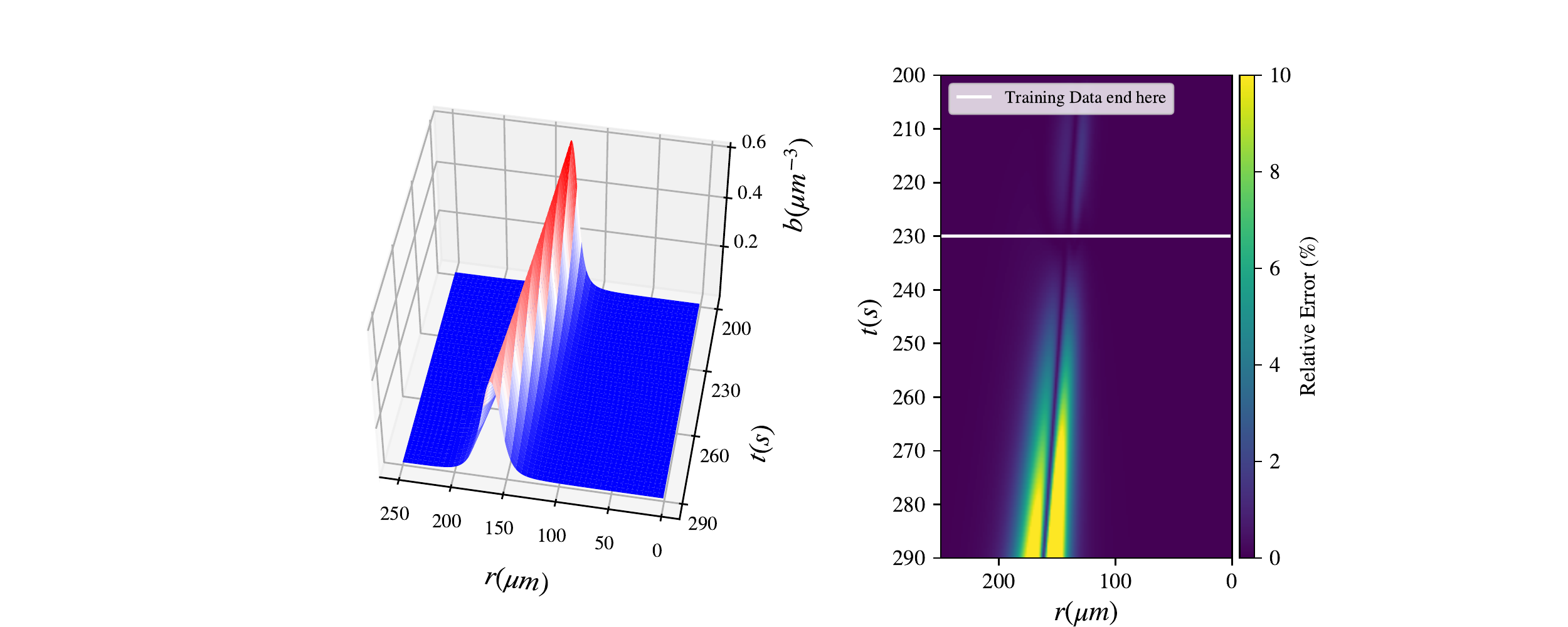}}
     \captionof{figure}{Black-box learning of both PDEs with Gaussian Process Regression: (left) Integration results for the \textbf{first} data-driven PDE  (for $b(r,t)$) and (right) relative error (\%).  Note that the white horizontal line separates the training dataset from the rest of the validation dataset.}
\label{fig:fig2GPBBB}
\end{center}

\begin{center} 
     \makebox[\textwidth][c]{\includegraphics[width=17.5cm, height=7cm]{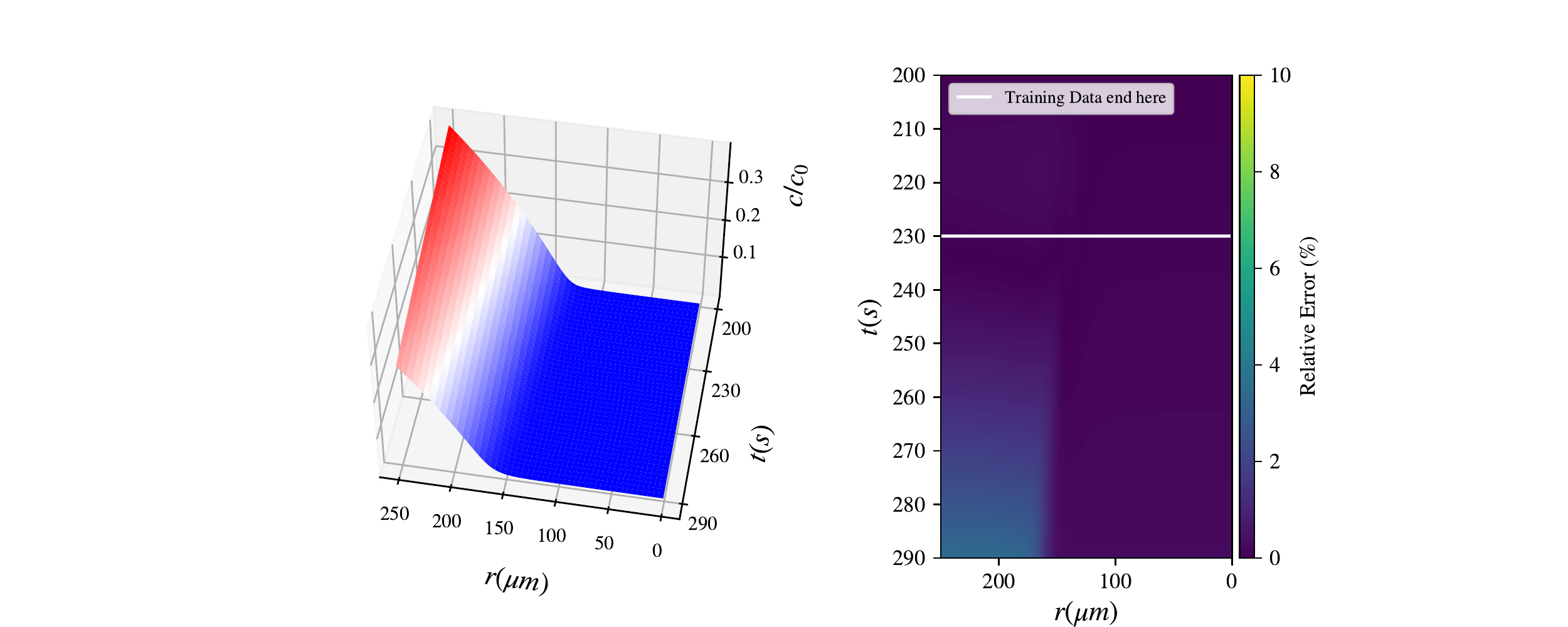}}
     \captionof{figure}{Black-box learning of both PDEs with Gaussian Process Regression: (left) Integration results for the \textbf{second} data-driven PDE (for $c(r,t)$) and (right) relative error (\%).}
\label{fig:fig2GPBBC}
\end{center}

\subsection{Black box learning of $b_t$ with GPR - $c_t$  known (with fields $b(r,t), c(r,t)$ known).}
\label{SubSubSecBBGP}

\begin{equation} \label{eqGPBB}
    b_t= f_{GP}(b, \mathbf{\nabla b} \cdot \mathbf{\hat{r}},\Delta b, c, \mathbf{\nabla c} \cdot \mathbf{\hat{r}},\Delta c )
\end{equation}

\begin{center}
     \makebox[\textwidth][c]{\includegraphics[width=20cm, height=8cm]{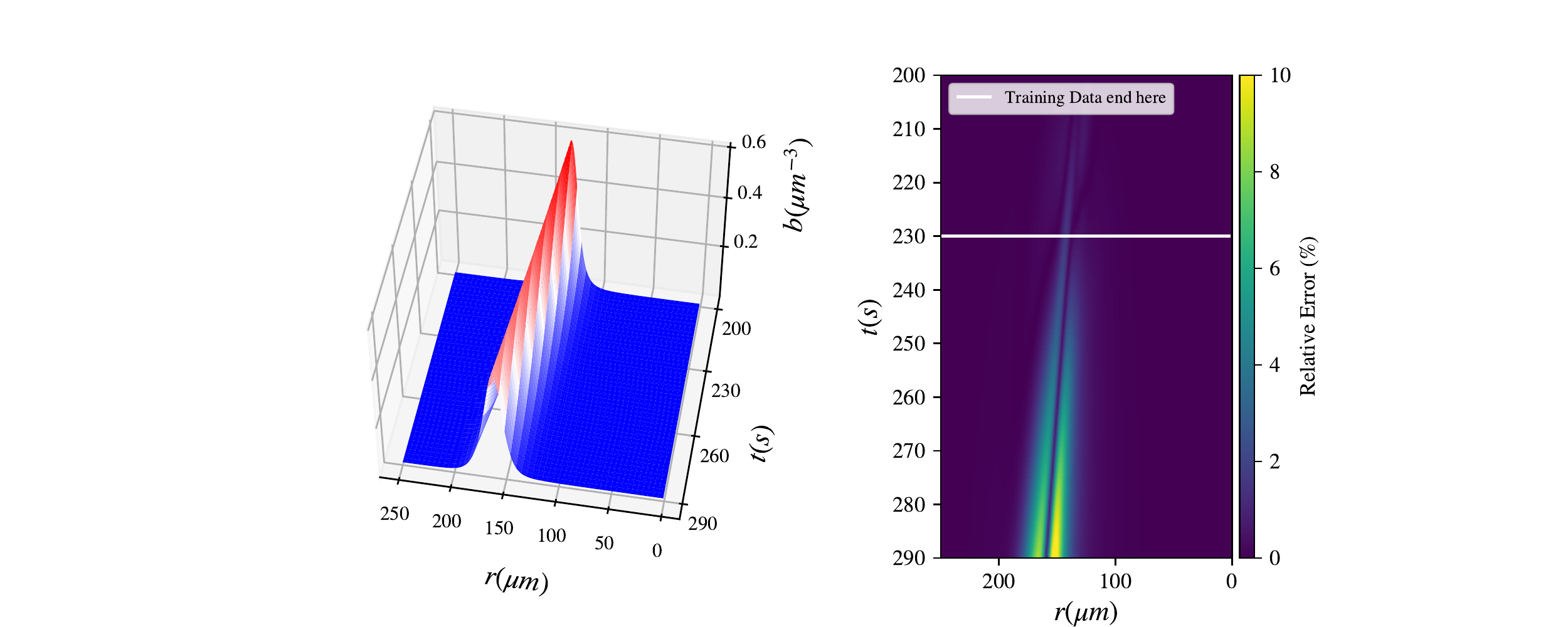}}
     \captionof{figure}{Black-box learning with Gaussian Process Regression: (left) Integration results for the data-driven PDE and (right) \% relative error.}
     \label{fig:figGPBB}
\end{center}

\subsection{Black box learning of $b_t$ - partial information with GPR (with only field $b(r,t)$ known)}
\label{SubSubSecBBGPpar}

\begin{equation} \label{eqGPBBpar}
    b(t_{k+1})=b(t_k)+ \Delta t f^{partial}_{GP}(b(t_k), (\mathbf{\nabla b} \cdot \mathbf{\hat{r}})(t_k),(\Delta b) (t_k), b(t_{k-1}), (\mathbf{\nabla b} \cdot \mathbf{\hat{r}})(t_{k-1}),(\Delta b)(t_{k-1}) ),
\end{equation}

with $\Delta t = t_{k+1} - t_k$, for any time point $t_k, k \geqslant 1$.

\begin{center}
     \makebox[\textwidth][c]{\includegraphics[width=17.5cm, height=7cm]{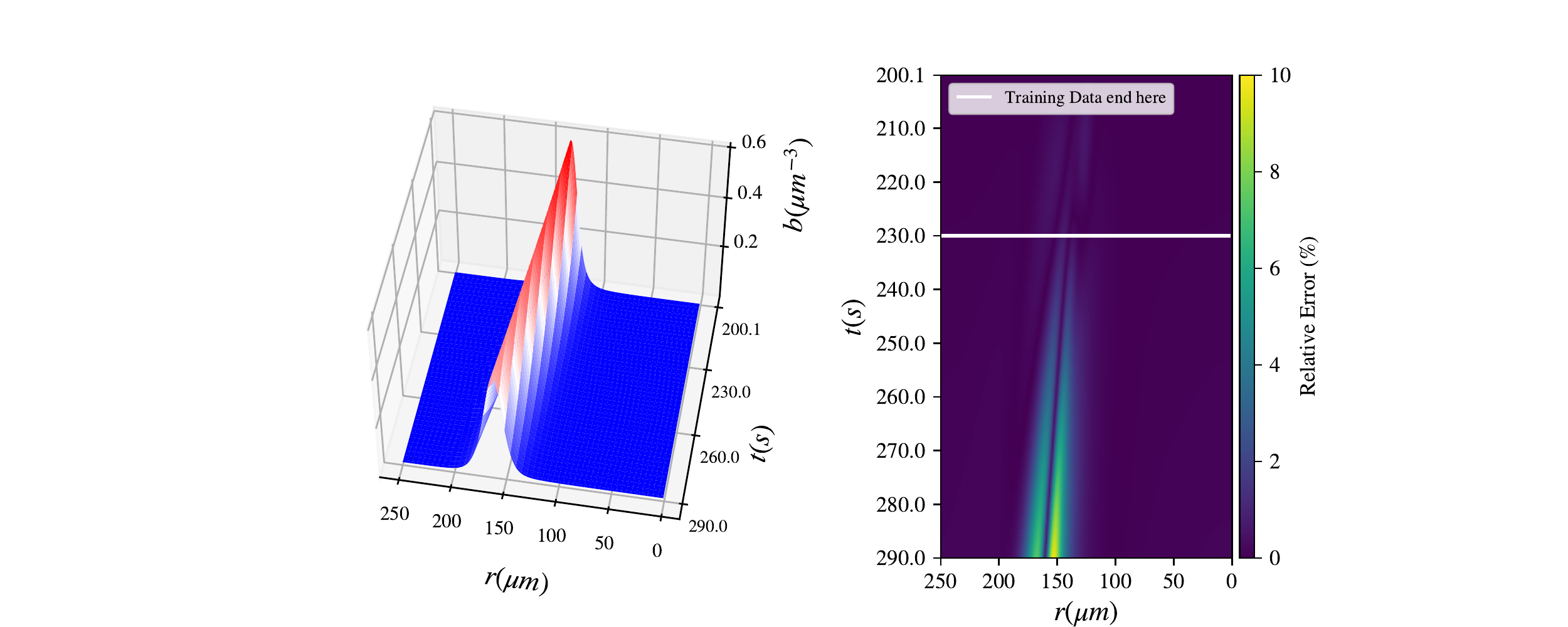}}
     \captionof{figure}{Black-box partial-information learning with Gaussian Process Regression: (left) Integration results for the data-driven PDE and (right)  \% relative error.}
     \label{fig:figGPBBpar}
\end{center}

\subsection{Gray box learning with GPR - $c_t$  known (with fields $b(r,t), c(r,t)$ known).}
\label{SubSubSecGBGP}

\begin{equation} \label{eqGPGB}
    b_t - D_b\Delta b = g_{GP}(b, \mathbf{\nabla b} \cdot \mathbf{\hat{r}},\Delta b, c, \mathbf{\nabla c} \cdot \mathbf{\hat{r}},\Delta c )
\end{equation}

\begin{center}
     \makebox[\textwidth][c]{\includegraphics[width=17.5cm, height=7cm]{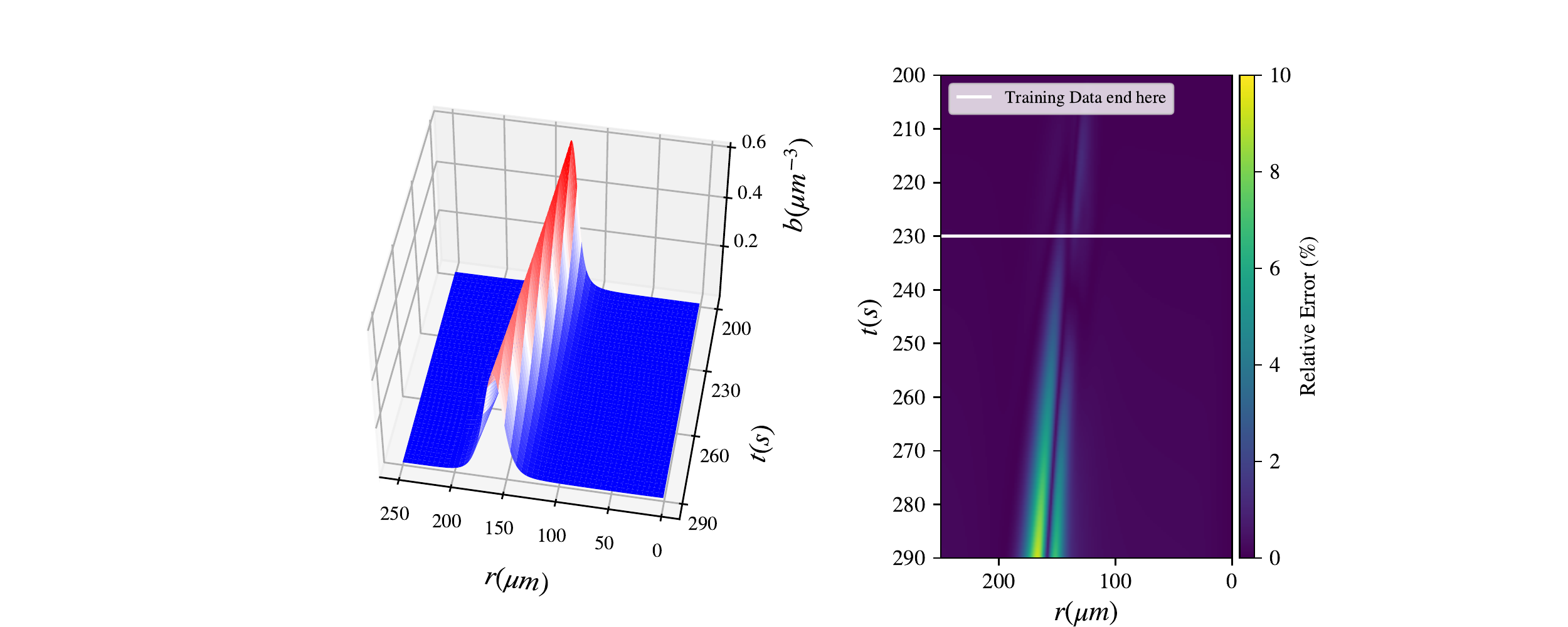}}
     \captionof{figure}{Gray-box learning with Gaussian Process Regression: (left) Integration results for the data-driven PDE and (right) \% relative error.}
     \label{fig:figGPGB}
\end{center}

\subsection{Gray box learning - partial information with GPR (with only field $b(r,t)$ known).}
\label{SubSubSecGBGPpar}

\begin{align} \label{eqGPGBpar}
    b(t_{k+1})=b(t_k) + \Delta t (D_b\Delta b(t_k)+ g^{partial}_{GP}( &b(t_k), (\mathbf{\nabla b} \cdot \mathbf{\hat{r}})(t_k),(\Delta b) (t_k),\\ \nonumber &b(t_{k-1}), (\mathbf{\nabla b} \cdot \mathbf{\hat{r}})(t_{k-1}),(\Delta b)(t_{k-1}) )),
\end{align}

with $\Delta t = t_{k+1} - t_k$, for any time point $t_k, k \geqslant 1$.

\begin{center}
     \makebox[\textwidth][c]{\includegraphics[width=20cm, height=8cm]{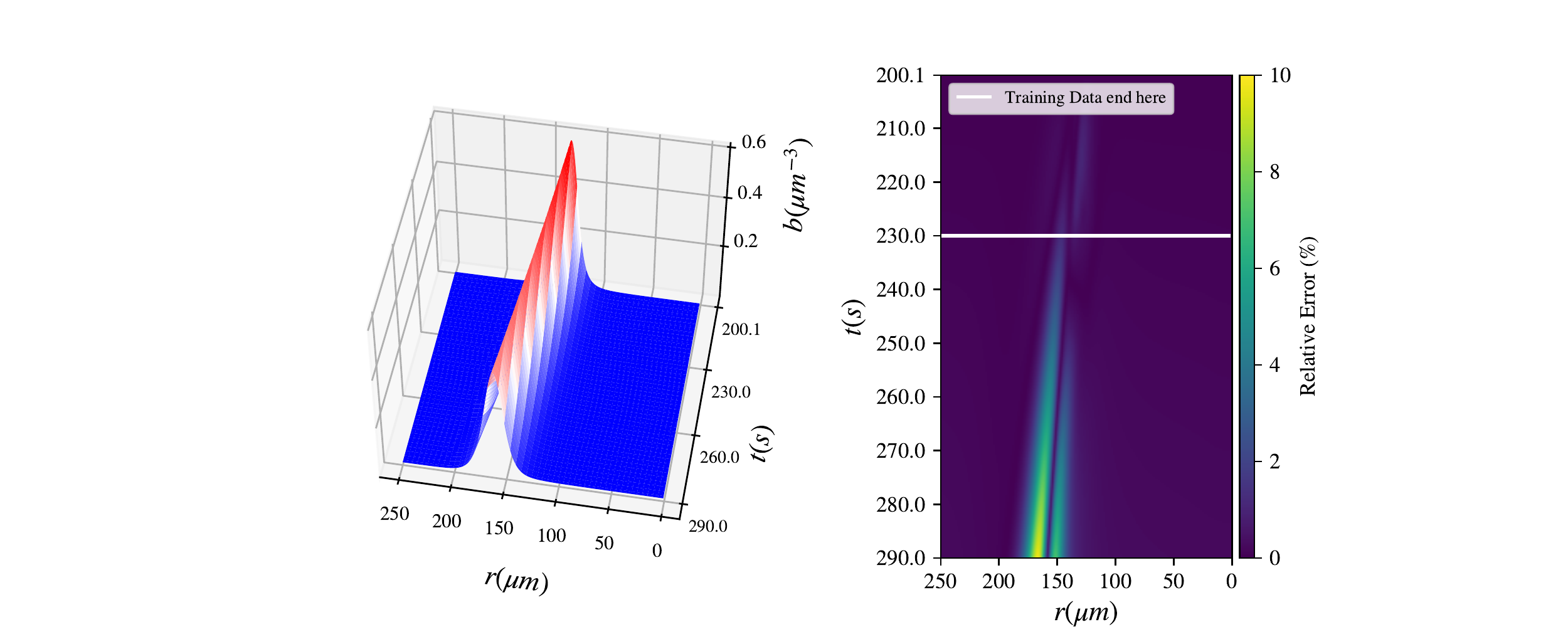}}
     \captionof{figure}{Gray-box partial-information learning with Gaussian Process Regression: (left) Integration results for the data-driven PDE and (right) \% relative error.}
     \label{fig:figGPGBpar}
\end{center}

\subsection{Gray box learning - partial information b with ANN (with only field $b(r,t)$ known).}
\label{SubSubSecGBNNpar}

\begin{align} \label{eqNNGBpar}
    b(t_{k+1})=b(t_k) + \Delta t (D_b\Delta b(t_k)+ g^{partial}_{NN}( &b(t_k), (\mathbf{\nabla b} \cdot \mathbf{\hat{r}})(t_k),(\Delta b) (t_k), \\ \nonumber &b(t_{k-1}), (\mathbf{\nabla b} \cdot \mathbf{\hat{r}})(t_{k-1}),(\Delta b)(t_{k-1}) )),
\end{align}

with $\Delta t = t_{k+1} - t_k$, for any time point $t_k, k \geqslant 1$.

\begin{center}
     \makebox[\textwidth][c]{\includegraphics[width=20cm, height=8cm]{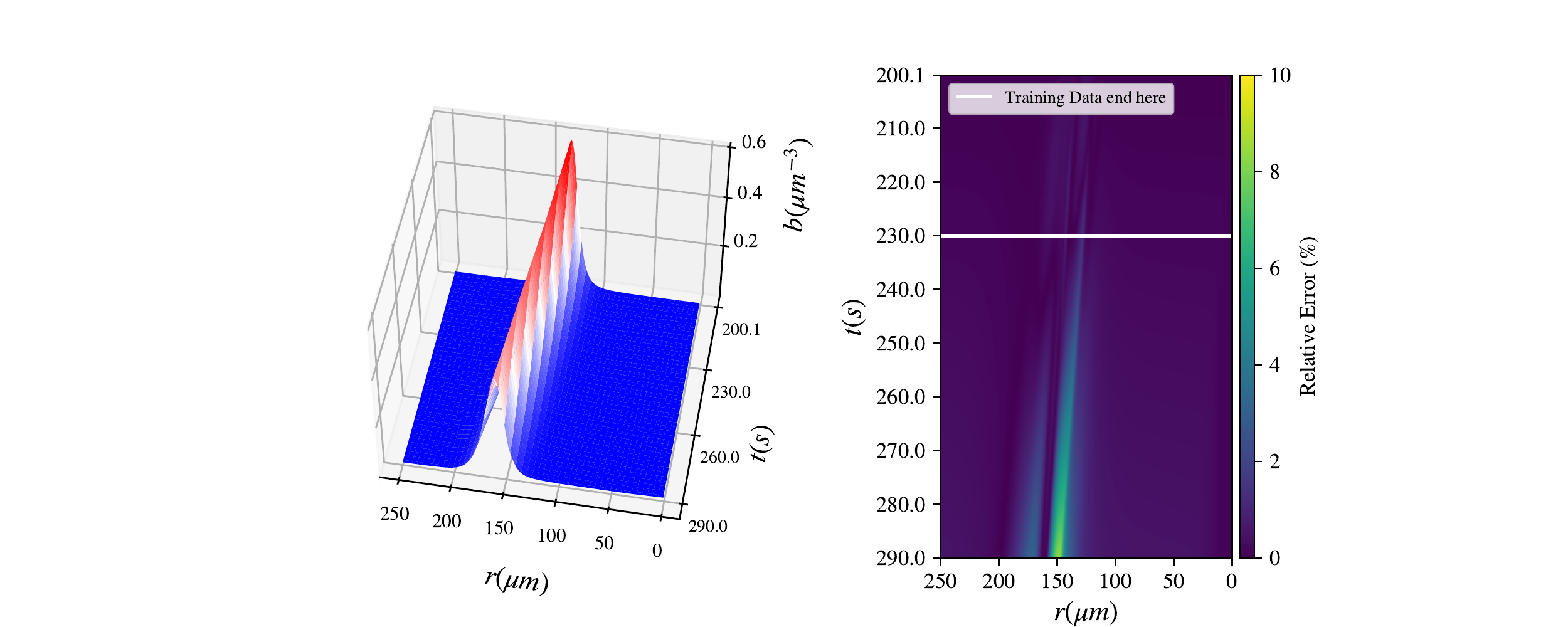}}
     \captionof{figure}{Gray-box partial-information learning with a Neural Network: (left) Integration results for the data-driven PDE and (right) \% relative error.}
     \label{fig:figNNGBpar}
\end{center}


\subsection{GPR - Learning c (with only field $b(r,t)$ known).}
\label{SubSubSecCGP}

\begin{align} \label{eqGPC}
    c(r_i, t_k) = C_{GP}(&b(r_i, t_k), \arctan\left(\frac{\overline{(\mathbf{\nabla b} \cdot    \nonumber
    \mathbf{\hat{r}})(r_i, t_k)}} {\overline{b(r_i, t_k)}}\right),(\Delta b) (r_i,t_k),\\  \nonumber
    & b(r_i, t_{k-1}), (\mathbf{\nabla b} \cdot \mathbf{\hat{r}})(r_i,t_{k-1}),(\Delta b)(r_i,t_{k-1})), 
\end{align}

for any discretization point is space $r_i$ and time point $t_k, k \geqslant 1$.

\begin{center}
     \makebox[\textwidth][c]{\includegraphics[width=20cm, height=8cm]{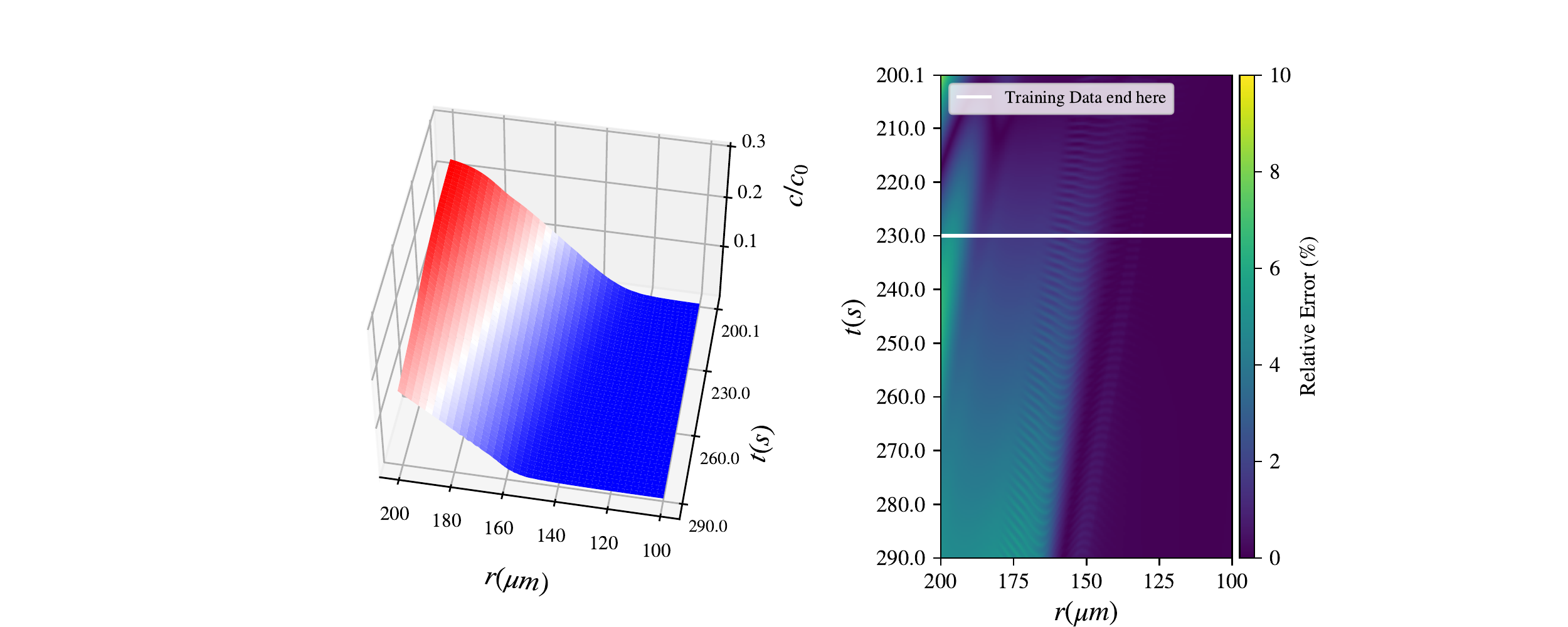}}
     \captionof{figure}{Learning the c-field with Gaussian Process Regression: (left) Field prediction and (right) \% relative error.}
     \label{fig:figGPc}
\end{center}

\section{Coordinate-independent operator learning}
\label{SecExt}

The gradient, curl, divergence, and Laplacian all arise by combining the exterior derivative $\d$, the metric tensor $g$, the exterior product $\wedge$ and the inner product of differential forms $\langle \cdot, \cdot \rangle$, the Hodge star operator $\star$, as well as the musical isomorphisms $^\sharp$ and $^\flat$.
Consequently, we can extend the problem of learning $f$ in
\begin{equation*}
  u = f(u, \grad u, \div \grad u, \dotsc)
\end{equation*}
to the more general setting of learning a function $f$ defined by compositions of the operators mentioned above.
For instance, the Navier-Stokes equations~\cite{gurtin1981} can be written in coordinate-free form as~\cite{wilson2011}
\begin{equation*}
  \left\{
    \begin{aligned}
      &\frac{\partial \omega}{\partial t} =  -\star (\omega \wedge \star \d \omega) - \nu \d \star \d \omega + \tfrac{1}{2} \d \langle \omega, \omega \rangle - \d p \\
      &\d \star \omega = 0,
    \end{aligned}
  \right.
\end{equation*}
where $\nu \ge 0$ is the viscosity and $p$ is the pressure.

The use of exterior calculus and exterior differential systems~\cite{bryant1991} in Physics-informed neural networks is currently growing~\cite{sitzmann2020,weiler2021,jenner2021,bronstein2021} and a more in-depth study of this framework is an interesting problem for future work.

\clearpage
\bibliography{Bibliography}